\documentclass{aa}  
\usepackage{graphicx}
\usepackage{txfonts}
\begin{document}

\title{Stellar populations of
early-type galaxies in different environments II}

\subtitle{Ages and metallicities}

\author{P. S\'{a}nchez--Bl\'{a}zquez\inst{1,2}
\and
J.\ Gorgas\inst{2}
\and
N.\ Cardiel\inst{2,3}
\and
J.J.\ Gonz\'alez\inst{4}} 
\institute{
Laboratoire d'Astrophysique, \'Ecole Polytechnique F\'ed\'erale de Lausanne (EPFL),
Observatoire, 1290 Sauverny, Switzerland\\
\and
Dpto.\ de Astrof\'{\i}sica, Fac. de Ciencias F\'{\i}sicas, Universidad
Complutense de Madrid, E-28040 Madrid, Spain \\
\and
Calar Alto Observatory, CAHA, Apartado 511, E-04004 Almer\'{\i}a,
Spain\\
\and
Instituto de Astronom\'{\i}a, Universidad Nacional Aut\'onoma de
M\'{e}xico, Apdo-Postal 70--264, M\'exico D.F, M\'exico}

\date{Accepted  . Received ; in original form
}

\abstract
{}
{This is the second paper of a series devoted to study the stellar 
content of early-type galaxies. The goal of the series is to set constraints on the 
evolutionary status of these objects}
{We use a new set of models which include an improved stellar library (MILES)
 to derive simple stellar population (SSP)-equivalent parameters in a sample of
 98 early-type galaxies. The sample contains galaxies  in the field, poor
 groups, and galaxies in the Virgo and Coma clusters. }
{We find that low-density environment   galaxies span a larger range in SSP age
and metallicity than their counterparts in high density environments, with a
tendency for lower $\sigma$ galaxies to be younger. 
Early-type galaxies in low-density environments appear on average $\sim$ 1.5 Gyr younger
and more metal rich than their counterparts in high density environments.  The
sample of low-density environment galaxies shows an age metallicity relation in
which younger galaxies are found to be more metal rich, but only when
metallicity is measured with a Fe-sensitive index. Conversely, there is no
age-metallicity relation when the metallicity is measured with a Mg sensitive
index.  The mass-metallicity relation is only appreciable for the low-density
environment galaxies when the metallicity is measured with a Mg-sensitive index
and not when the metallicity is measured with other indicators. On the
contrary, this relation exists for the high-density environment galaxies
independently of the indicator used to measure the metallicity. }
{This suggests a
dependence of the mass-metallicity relation on the environment of the galaxies.
Our data favour a scenario in which galaxies in low density environments have
suffered a more extended star formation history than the galaxies in the Coma
cluster, which appear to host more homogenous stellar populations.}

\maketitle

\section{Introduction}

The knowledge of the star formation history of early-type galaxies is a key
test of our  understanding of the galaxy formation processes. The classic
vision of elliptical galaxies as old objects forming their stars at high
redshift in a single episode has come into question by several studies of
stellar population in these systems which found a high fraction of early-type
galaxies with apparent young ages (e.g. Gonz\'alez 1993; Trager et al.\ 1998;
Trager et al.\ 2000b; Terlevich \& Forbes 2002; S\'anchez--Bl\'azquez 2004).


The semi-analytical models of galaxy formation (Kauffmann 1996; Baugh et al.
1996; Cole et al.\ 1994; Somerville et al.\ 1999; de Lucia et al.\ 2006), in 
the framework of  cold dark matter (CDM) cosmological model and in the current
standard $\Lambda$CDM, 
predict extended star formation histories for early-type galaxies,
with substantial fractions of the stellar population formed at relative low
redshift, in agreement with the observed trends.  The numerical simulations
based on the semi-analytical models can reproduce impressively various features
of large-scale structures from dwarf galaxies to  giant galaxies and rich
clusters of galaxies (e.g. Steinmetz \& Navarro 2002; Klypin et al 2003).
One of the keys to test the hierarchical scenarios is to study the
properties of galaxies situated in different environments, since the
semi-analytical models predict that galaxies in dense clusters were assembled
at higher redshift than galaxies in the field and poor groups.

This is the second paper of a series devoted to the study of the stellar
content in nearby early-type galaxies. The final aim of the series is to
constrain the formation epoch of the stellar population in this kind of
galaxies as a function of mass and environment.  The first paper of the series
(S\'anchez--Bl\'azquez et al.\ 2006a, hereafter Paper I) analysed the relation
of the central line-strength indices with the velocity dispersion for a sample
of 98 galaxies drawn from different environments. 
It also presented  some
evidences of differences between the stellar content of galaxies in different
environments. In particular, we found that the index--$\sigma$ relations are
driven by both age and metallicity in the sample of galaxies in low-density
environments and in the Virgo cluster.
However, an age variation with $\sigma$ is not required to explain the
index--$\sigma$ relations for galaxies belonging to the Coma cluster.  We also
presented evidences supporting that the [Mg/Fe], [N/Fe], and probably [C/Fe]
ratios increase with the velocity dispersion of the galaxy in both subsamples.
In Paper I we studied the scatter in the index--$\sigma$ relations, finding
that this is not only a consequence of a dispersion in the age of the galaxies,
but it is also due to variations of the [Mg/Fe] ratio. These variations are
related to the mean ages of the galaxies, in the sense that younger galaxies
exhibit, on average, lower [Mg/Fe] ratios. Furthermore, galaxies in the Coma cluster 
show, on average, higher [Mg/Fe] ratios than galaxies in lower 
environment. We also detected systematic differences in the values of
some indices which were interpreted as differences in chemical abundances
ratios between both subsamples. 
   
In Paper I we analysed the raw line-strength indices and their relation with
other parameters.   In this paper, we compare these indices with the
predictions of the stellar population synthesis models by Vazdekis et al.\
(2006, in preparation; hereafter V06) to derive central simple stellar
population (SSP) parameters for our sample of galaxies. 
 
The outline of this paper is as follows. 
Section 2 details the estimation of age and metallicities. In Section 3 we
study the mean values of these parameters and their differences as a function
of the environment. In Section 4 we analyse the most likely scenarios to
explain the dispersion in the ages of the galaxies.  Section 5 is devoted to
present and discuss the age--metalicity relation for the galaxies in both
environments, and in Section 6 we study the relation of the age and
metallicity with the velocity dispersion. In Section 7 we present a brief
discussion of the results and Section 8 summarises our findings and
conclusions. 

The sample consists of 98 galaxies, out of which 35 belong to the rich cluster
of Coma (high-density environment galaxies, hereafter HDEGs) and the rest are
galaxies in the field, in groups or in the Virgo cluster (low-density
environment galaxies, hereafter LDEGs).  For a detailed description of the
sample we refer the reader to Paper I. 

\section{Derivation of ages and metallicities}
\label{derivation_age_metallicity}
 
Previous works have used Lick/IDS line-strength indices to derive mean ages and
metallicities using evolutionary synthesis models (e.g. Worthey 1994; Buzzoni
1995; Bruzual \& Charlot 2003; Thomas, Maraston \& Bender 2003).  Here we
follow a similar approach, deriving the SSP parameters  (age and metallicity)
by comparing the observed line-strengths with the predicted index--index
diagrams from a new set of models by V06.  These models are an updated 
version of those described by Vazdekis et al.\ (2003) improved by the inclusion 
of  
a  new stellar library (MILES) recently  observed by
S\'anchez--Bl\'azquez et al.\ (2006). This library contains 1003 stars,
carefully selected to cover the atmospheric parameter space in an homogenous
way. In particular, the library  span a range of metallicities from [Fe/H]$\sim
-2.7$ to $+1$, and a wide range of effective temperatures.  The inclusion of
this library reduces the uncertainties in the models, specially at
metallicities departing from solar.  Since the stars of the library are
relatively flux calibrated, these models are able to predict, not only
individual features for a population of a given age and metallicity, but the
whole spectral energy distribution (SED).  This allows to analyse the spectra
of the galaxies at its own resolution, given by their internal velocity and
instrumental broadening (see e.g., Vazdekis et al.\ 2001).  The synthetic 
spectra have a spectral resolution of 2.3 \AA  and  cover the spectral range 
 3500-7500\AA.  
 
In spite of this capability,  as we are using calibrations based on the Lick
system, and in order to compare our results with previous
studies, most of the analysis has been performed with the indices transformed
into the Lick system. Therefore, in order to compare with the model
predictions, we broadened the synthetic spectra to match the wavelength
dependent resolution of the Lick stellar library (Worthey \& Ottaviani 1997)
and measured the indices in the same way as in the galactic spectra. We then
added to the synthetic indices the same offsets that we applied to the galaxy
indices. (see Paper I for details).

\begin{figure*}
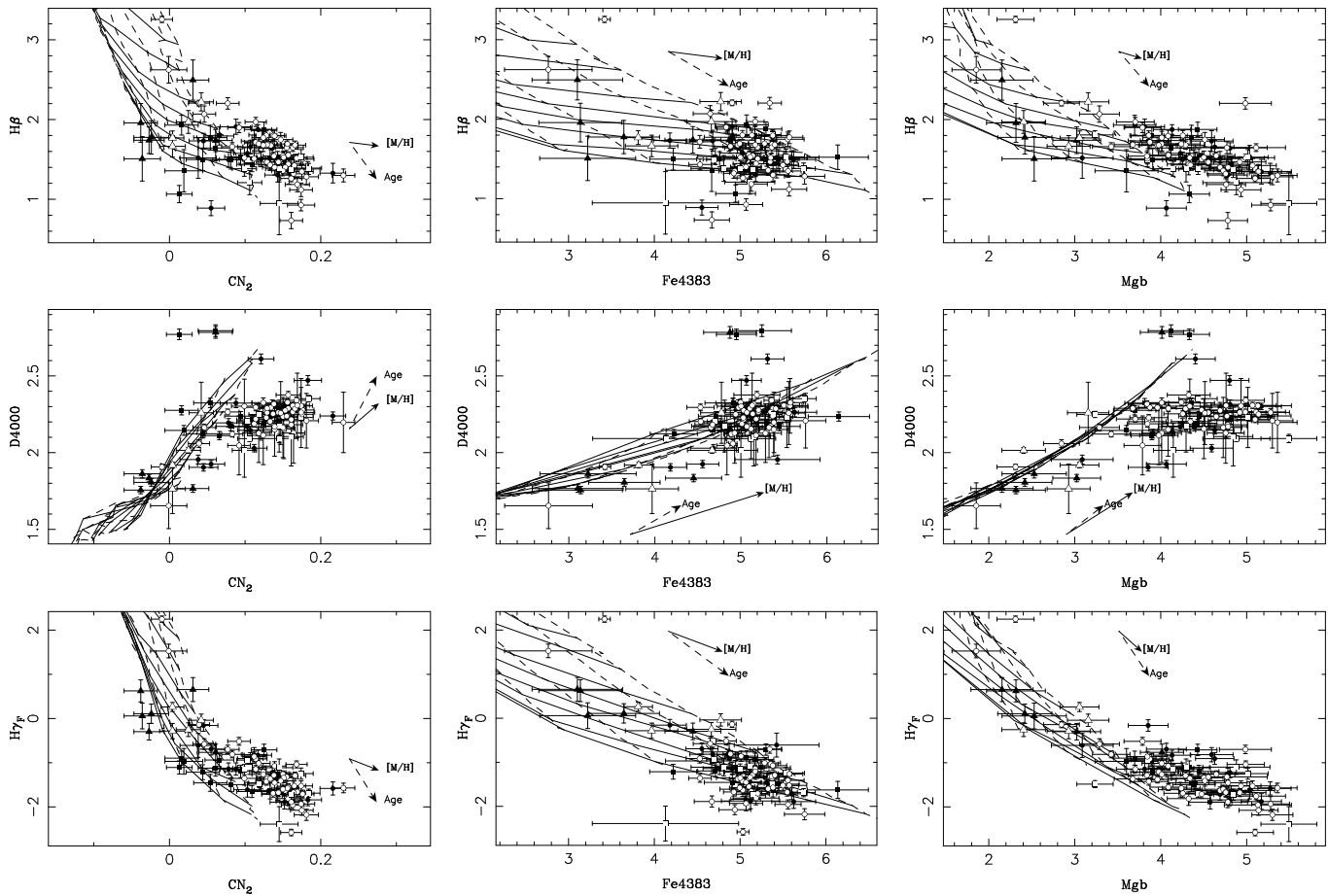

   \resizebox{0.32\hsize}{!}{\includegraphics[angle=-90]{cn2.hbeta.paper.ps}}
   \hfill
   \resizebox{0.32\hsize}{!}{\includegraphics[angle=-90]{fe4383.hbeta.paper.ps}}
   \hfill
   \resizebox{0.32\hsize}{!}{\includegraphics[angle=-90]{mgb.hbeta.paper.ps}} 

   \vspace{2mm}

   \resizebox{0.32\hsize}{!}{\includegraphics[angle=-90]{cn2.d4000.paper.ps}}
   \hfill
   \resizebox{0.32\hsize}{!}{\includegraphics[angle=-90]{fe4383.d4000.paper.ps}}
   \hfill
   \resizebox{0.32\hsize}{!}{\includegraphics[angle=-90]{mgb.d4000.paper.ps}}

   \vspace{2mm}

   \resizebox{0.32\hsize}{!}{\includegraphics[angle=-90]{cn2.hgf.paper.ps}}
   \hfill
   \resizebox{0.32\hsize}{!}{\includegraphics[angle=-90]{fe4383.hgf.paper.ps}}
   \hfill
   \resizebox{0.32\hsize}{!}{\includegraphics[angle=-90]{mgb.hgf.paper.ps}}        
 \caption{Line strengths of LDEGs (open symbols) and HDEGs (filled symbols)
 through  the central aperture.  The distinct symbols indicate different kind
 of galaxies: triangles correspond to dwarf ellipticals, squares to
 lenticulars, and circles to normal ellipticals.  Model grids from V06 have
 been superposed: solid lines are contours of constant age (1.0, 1.41, 2.00,
 2.82, 3.98, 5.62, 7.98, 11.22, 15.85  Gyr) and dashed lines are contours of
 constant [M/H] ([M/H]=$-0.68$, $-0.38$, 0.0, +0.2 dex). The arrows indicate
 the direction of increasing age and metallicity.  Three different metal
 indicators (CN$_2$, Fe4383 and Mgb) are combined with three different
 age-sensitive index (H$\beta$,  D4000 and H$\gamma_{\rm F}$). 
 \label{index-index}} 
\end{figure*}

Fig. \ref{index-index} shows several index--index diagrams combining different
pairs of indices.  Over-plotted are the stellar population models of V06 for
various ages and metallicities as indicated in the figure caption.  Open and
filled symbols represent LDEGs and HDEGs respectively.  It is clear from the
figure that galaxies span a fair range in their mean ages. This result has
been previously found by other authors (Gonz\'alez 1993; Trager et al.\ 1998;
Trager et al.\ 2000a) and it is in contradiction to the classical vision of
early-type galaxies as old and coeval systems.

To quantify the age and metallicity values, we interpolated in the grids using
bivariate polynomials, as described in Cardiel et al.\ (2003).
Although the highest metallicity modeled by V06 is $[{\rm M}/{\rm H}]=+0.2$,
in order to obtain metallicity  values for the maximum number of galaxies, we
extrapolated the predictions up to $[{\rm M}/{\rm H}]=+0.5$.  However, all the
values above $[{\rm M}/{\rm H}]=+0.2$ have to be considered more uncertain.
The errors in age and metallicity were calculated as the differences between
the central values and the values at the end of the error bars, which give an
upper limit to the real errors.  Table \ref{edades.metalicidades} lists the
ages and metallicities derived from  several index--index diagrams, all of them
combining H$\beta$ with other metal-sensitive indices.  The empty spaces in the
table indicate that the galaxy lies outside the model grid, and therefore no
measurements of age and metallicity have been made.  The ages and metallicities
presented here represent  simple stellar population equivalent parameters, and
thus, if all the stars were not formed in a single event,
they represent values weighted with the luminosity of the stars and do not
necessarily reflect the age and the metallicity of the bulk of the stars  in
the galaxy. We have used H$\beta$ as the main age indicator as the other
higher-order Balmer lines (H$\delta$ and H$\gamma$) are very sensitive to
$\alpha$/Fe ratio changes at super-solar metallicities (Thomas, Maraston \&
Korn 2004).

\subsection{The problem of the relative abundances}
 One aspect which is evident in Fig. \ref{index-index} is that the ages
and metallicities obtained with different indicators are not the same. 
In particular, the metallicities measured with indices such as CN$_2$, Mgb and
C4668 are larger than the metallicities measured with  Fe-sensitive indices
such as Fe4383.
This effect has been noted previously by several authors (eg. Peletier 1989; 
Faber et al. 1999; Vazdekis et al.\ 2001; Thomas, Maraston \& Korn 2004; among
others) and is commonly attributed 
to an overabundance of Mg (O'Connell 1976), N (Worthey 1998) and C (Vazdekis et
al.\ 2001) with respect to Fe\footnote{Although we use the term overabundance,
in fact Trager et al. (2000b) has shown that the differences are more likely
due to a depression of Fe with respect to the solar values more than to an
enhancement of the light elements.} compared to the solar abundance partition.   
The differences in the relative chemical abundance ratios with respect to the
solar values is one of the major roadblocks to derive reliable stellar
population parameters.  However, since different elements are produced in
stars with different lifetimes, the study of the chemical composition of
early-type galaxies is a powerful tool to disentangle the star formation
history of these systems. Unfortunately, the computation of chemical abundances
through the comparison with  stellar population models is still in its infancy.
Great efforts have been made  to include the non-solar partition effects in the
theoretical models through the building of $\alpha$-enhanced\footnote{The so
called $\alpha$-elements: O, Ne, Mg, Si, S, Ar, Ca, Ti, are particles build up
with $\alpha$-particle nuclei.} isochrones (Salaris, Chieffi \& Straniero 1993;
Salasnich et al.\ 2000) and stellar model atmospheres with variations on the
abundance of individual elements (Tripicco \& Bell 1995). Some attempts have
also been made to include these theoretical work into the SSP models
(Trager et al.\ 2000b; Thomas et al.\ 2003). But, despite all these
valuable works,  we are still lacking  models which include the effect of
variations of chemical abundances ratios in a consistent way (with isochrones,
model atmospheres and  stellar libraries). The current models have an
inconsistency between the atmospheres and the stellar interiors.

The potential effects of including isochrones with non-solar chemical abundance
ratios is not clear. Tantalo, Chiosi \& Bressan (1998) built a set of stellar
isochrones with relative abundances of [$\alpha$/Fe] different than solar and
concluded that the models incorporating those were indistinguishable in the
colour-magnitude diagram from the models with the same metallicity and
$[\alpha/$Fe$]=0$. Salaris et al.\ (1993) showed (only for metallicities lower
than solar) that the isochrones with overabundances of $\alpha$ elements were
identical to the ones scaled to solar ratios at the same metallicity, if the
ratio between the mass fraction of elements with high- and low- ionisation
potential were constant.  However, more recently, Salaris \& Weiss (1998) have
suggested that at higher metallicities (near solar), the isochrones shift to
higher temperatures and their shapes change as the ratio $\alpha$/Fe increases.
Trager et al.\ (2001b) have analysed the effect of this change and have not
found much difference in the predicted indices. However, Thomas \& Maraston
(2003) found large differences in the inferred ages using models which
incorporate the isochrones with $\alpha$-enhancement with respect to the models
which do not. So far, the effect of including isochrones with [C/Fe] and
[N/Fe] different from solar has not been studied. 
 
Another problem is that the empirical stellar libraries included in the
population models are limited to stars in the solar neighborhood with,
therefore, relative compositions between different elements resembling the
solar one. The ratios between the different elements in these stars are not
well known but, assuming that they follow the trends  of the disk stars in the
Galaxy, these abundances are not constant with metallicity, which, if  not
taken properly into account, can produce  an apparent variation of the relative
abundances with metallicity (Proctor et al.\ 2004).

For all these reasons, in this work we do not attempt to derive relative
abundances of the different elements. However, with the aim of exploring the
behaviour of the different chemical species, we measure the metallicity using
several indices especially sensitive to variations of different elements.

For the rest of the analysis we make the assumption that the differences
between the metallicities derived from various index--index diagrams, combining
H$\beta$ with other metallicity indicators, are due to changes in the
sensitivity of these indicators to variations of different chemical abundances. 

\begin{table*}
\caption{Central ages and metallicities obtained in different index--index
diagrams. The associated errors are indicated under the measurements. See text
section \ref{derivation_age_metallicity}) for more details.\label{edades.metalicidades}}
\centering
\begin{tabular}{@{}l  r  r r r r r r r r r@{}}
\hline\hline
Galaxy      & $\log$(age)&   [M/H]&  $\log$(age)&[M/H]&$\log$(age)&[M/H]&$\log$(age)&[M/H]&$\log$(age)&   [M/H]\\
              &\multicolumn{2}{c}{[MgbFe]--H$\beta$}&
              \multicolumn{2}{c}{Fe4383--H$\beta$}&\multicolumn{2}{c}{Mgb--H$\beta$}&\multicolumn{2}{c}{CN$_2$--H$\beta$}&
               \multicolumn{2}{c}{(spectral synthesis)}\\
\hline
NGC 221       &9.455 &0.164   &   9.431&$   0.283$&    9.464&   0.105&     9.443&   0.223& 9.698&$ -0.003$\\
              &0.029 &0.087   &   0.021&$   0.028$&    0.036&   0.150&     0.263&   0.202& 0.066&$  0.053$\\
NGC 315       &9.913 &0.517   &         &         &    9.913&   0.517&     9.931&   0.411&10.066&$  0.079$\\
              &0.028 &0.053   &         &         &    0.028&   0.053&     0.032&   0.036& 0.039&$  0.031$\\
NGC 507       &9.996 &0.196   &  10.186&$  -0.180$&    9.945&   0.311&     9.938&   0.342&10.078&$  0.093$\\
              &0.091 &0.129   &   0.169&$   0.223$&    0.077&   0.150&     0.105&   0.091& 0.038&$  0.032$\\
NGC 584       &9.748 &0.213   &   9.828&$   0.120$&    9.719&   0.333&     9.704&   0.403& 9.790&$  0.125$\\
              &0.110 &0.085   &   0.094&$   0.095$&    0.162&   0.125&     0.161&   0.111& 0.046&$  0.031$\\
NGC 636       &9.768 &0.207   &   9.859&$   0.082$&    9.727&   0.316&     9.703&   0.421& 9.701&$  0.110$\\
              &0.104 &0.082   &   0.089&$   0.092$&    0.162&   0.126&     0.160&   0.107& 0.042&$  0.032$\\
NGC 821       &9.714 &0.299   &   9.733&$   0.215$&    9.571&   0.502&     9.568&   0.527&10.063&$  0.113$\\
              &0.180 &0.183   &   0.138&$   0.084$&    0.163&   0.086&     0.159&   0.195& 0.044&$  0.032$\\
NGC 1600      &9.935 &0.300   &  10.075&$   0.014$&         &        &     9.807&   0.442&10.094&$  0.120$\\
              &0.052 &0.060   &   0.113&$   0.112$&         &        &     0.120&   0.042& 0.035&$  0.032$\\
NGC 1700      &9.706 &0.283   &   9.725&$   0.193$&    9.579&   0.405&     9.561&   0.517&9.833 &$  0.131$\\
              &0.076 &0.071   &   0.095&$   0.073$&    0.097&   0.089&     0.031&   0.196&0.046 &$  0.030$\\
NGC 2300      &9.905 &0.289   &   9.943&$   0.101$&         &        &     9.869&   0.438&9.927 &$  0.136$\\
              &0.100 &0.083   &   0.129&$   0.118$&         &        &     0.062&   0.148&0.046 &$  0.030$ \\
NGC 2329      &10.153&0.127   &         &        &    9.955&   0.445&    10.122&   0.168 &9.923 &$  0.060$\\
              &0.155 &0.197   &        &        &    0.158&   0.129&     0.128&   0.134  &0.061 &$  0.053$ \\
NGC 2693      &10.173&0.197   &        &        &    9.988&   0.435&    10.027&   0.357  &9.988 &$  0.131$\\
              &0.114 &0.084   &        &        &    0.086&   0.094&     0.064&   0.060  &0.044 &$  0.030$\\
NGC 2694      &9.931 &0.162   &  10.029&$  -0.059$&    9.638&   0.399&     9.920&   0.212&9.726 &$  0.165$\\
              &0.195 &0.158   &   0.155&$   0.152$&    0.273&   0.130&     0.155&   0.161&0.040 &$  0.024$\\
NGC 2778      &9.704 &0.288   &   9.737&$   0.134$&    9.569&   0.452&     9.561&   0.512&10.016&$  0.100$\\
              &0.071 &0.061   &   0.051&$   0.060$&    0.089&   0.076&     0.029&   0.307&0.039 &$  0.032$\\
NGC 2832      &9.949 &0.334   &  10.138&$  -0.027$&    9.913&   0.515&     9.764&   0.474&9.988 &$  0.131$\\
              &0.075 &0.098   &   0.101&$   0.138$&    0.035&   0.085&     0.229&   0.086&0.045 &$  0.031$\\
NGC 3115      &9.923 &0.297   &  10.050&$  -0.002$&    9.906&   0.402&     9.752&   0.399&9.931 &$  0.133$\\
              &0.025 &0.056   &   0.066&$   0.057$&    0.029&   0.071&     0.135&   0.091&0.046 &$  0.031$\\
NGC 3377      &9.715 &0.084   &   9.712&$   0.097$&    9.562&   0.341&     9.560&   0.359&9.950 &$ -0.010$\\
              &0.089 &0.097   &   0.081&$   0.071$&    0.102&   0.092&     0.102&   0.112&0.071 &$  0.068$\\
NGC 3379      &9.914 &0.172   &   9.974&$  -0.024$&    9.894&   0.287&     9.865&   0.405&9.935 &$  0.129$\\
              &0.033 &0.059   &   0.059&$   0.046$&    0.104&   0.109&     0.119&   0.163&0.046 &$  0.031$\\

\hline
\end{tabular}
\end{table*}
\begin{table*}
\addtocounter{table}{-1}
\caption{\it Continued.}
\centering
\begin{tabular}{@{}l  r  r r r r r r r r r@{}}
\hline\hline
Galaxy       & $\log$(age)&   [M/H]&  $\log$(age)&[M/H]&$\log$(age)&[M/H]&$\log$(age)&[M/H]&$\log$(age)&   [M/H]\\
              &\multicolumn{2}{c}{[MgbFe]--H$\beta$}&
              \multicolumn{2}{c}{Fe4383--H$\beta$}&\multicolumn{2}{c}{Mgb--H$\beta$}&\multicolumn{2}{c}{CN$_2$--H$\beta$}&
              \multicolumn{2}{c}{(spectral synthesis)}\\
\hline
NGC 3605      &9.512 &0.227   &   9.521&$   0.144$&    9.514&   0.210&          &        &9.925 &$  0.012$\\
              &0.142 &0.187   &   0.145&$   0.119$&    0.182&   0.227&          &        &0.071 &$  0.064$\\
NGC 3608      &9.926 &0.176   &  10.011&$  -0.033$&    9.878&   0.409&     9.829&   0.407&9.921 &$  0.122$\\
              &0.078 &0.085   &   0.105&$   0.097$&    0.129&   0.107&     0.125&   0.078&0.040 &$  0.031$\\
NGC 3641      &9.742 &0.216   &   9.870&$  -0.079$&    9.723&   0.292&     9.702&   0.392&10.105&$  0.107$\\
              &0.099 &0.083   &   0.087&$   0.075$&    0.153&   0.131&     0.150&   0.107&0.034 &$  0.032$\\
NGC 3665      &10.105&0.014   &  10.159&$  -0.131$&    9.967&   0.251&    10.003&   0.181& 9.910&$  0.045$\\
              &0.132 &0.188   &   0.087&$   0.100$&    0.152&   0.153&     0.108&   0.091& 0.065&$  0.052$\\
NGC 3818      &9.761 &0.218   &   9.870&$   0.018$&    9.729&   0.310&     9.597&   0.444&10.077&$  0.099$\\
              &0.090 &0.071   &   0.055&$   0.059$&    0.147&   0.117&     0.083&   0.065& 0.039&$  0.033$\\
NGC 4261      &9.709 &0.423   &   9.879&$   0.075$&         &        &     9.582&   0.544&10.095&$  0.118$\\
              &0.141 &0.069   &   0.047&$   0.053$&         &        &     0.140&   0.316& 0.034&$  0.032$\\
NGC 4278      &10.095&0.152   &        &        &    9.931&   0.500&         9.968&   0.372&10.066&$  0.079$\\
              & 0.080&0.072   &        &        &    0.063&   0.042&         0.038&   0.050& 0.039&$  0.031$\\
NGC 4365      & 9.900&0.344   &   9.978&$   0.040$&         &        &       9.881&   0.414&10.100&$  0.139$\\
              & 0.084&0.065   &   0.074&$   0.064$&         &        &       0.065&   0.139& 0.035&$  0.030$\\
NGC 4374      &10.054&0.118   &  10.148&$  -0.069$&    9.925&$   0.429$&     9.941&   0.354&10.017&$  0.099$\\
              & 0.103&0.105   &   0.045&$   0.052$&    0.029&$   0.062$&     0.083&   0.075& 0.039&$  0.032$\\
NGC 4415      & 9.855&-0.146  &   9.893&$  -0.256$&    9.841&$  -0.103$&          &        & 9.751&$ -0.424$\\
              & 0.104& 0.118  &   0.092&$   0.108$&    0.126&$   0.197$&          &        & 0.062&$  0.079$\\
NGC 4431      &10.001&-0.408  &   9.972&$  -0.264$&    9.973&$  -0.270$&          &        & 9.955&$ -0.483$\\
              & 0.125& 0.243  &   0.179&$   0.174$&    0.146&$   0.325$&          &        & 0.067&$  0.069$\\
NGC 4464      &10.057&-0.060  &  10.096&$  -0.154$&    9.943&$   0.164$&     9.929&   0.223& 9.846&$  0.115$\\
              & 0.085& 0.100  &   0.058&$   0.054$&    0.113&$   0.147$&     0.038&   0.056& 0.046&$  0.032$\\
NGC 4467      & 9.993& 0.027  &  10.043&$  -0.063$&    9.934&$   0.171$&     9.917&   0.254& 9.851&$  0.116$\\
              & 0.130& 0.135  &   0.122&$   0.113$&    0.156&$   0.180$&     0.119&   0.134& 0.046&$  0.031$\\
NGC 4472      & 9.984& 0.220  &  10.072&$   0.068$&    9.935&$   0.355$&     9.921&   0.429&10.098&$  0.139$\\
              & 0.084& 0.078  &   0.098&$   0.094$&    0.056&$   0.090$&     0.077&   0.050& 0.035&$  0.030$\\
NGC 4478      & 9.874& 0.055  &   9.885&$  -0.020$&    9.878&$   0.029$&     9.851&   0.199& 9.699&$  0.093$\\
              & 0.067& 0.088  &   0.049&$   0.051$&    0.109&$   0.168$&     0.081&   0.111& 0.044&$  0.034$\\
NGC 3379      & 9.914& 0.172  &   9.974&$  -0.024$&    9.894&   0.287&     9.865&     0.405& 9.935&$  0.129$\\
              & 0.033& 0.059  &   0.059&$   0.046$&    0.104&   0.109&     0.119&     0.163& 0.046&$  0.031$\\
NGC 3605      & 9.512& 0.227  &   9.521&$   0.144$&    9.514&   0.210&          &          & 9.925&$  0.012$\\
              & 0.142& 0.187  &   0.145&$   0.119$&    0.182&   0.227&          &          & 0.071&$  0.064$\\
NGC 3608      & 9.926& 0.176  &  10.011&$  -0.033$&    9.878&   0.409&     9.829&     0.407& 9.921&$  0.122$\\
              & 0.078& 0.085  &   0.105&$   0.097$&    0.129&   0.107&     0.125&     0.078& 0.040&$  0.031$\\
NGC 3641      & 9.742& 0.216  &   9.870&$  -0.079$&    9.723&   0.292&     9.702&     0.392&10.105&$  0.107$\\
              & 0.099& 0.083  &   0.087&$   0.075$&    0.153&   0.131&     0.150&     0.107& 0.034&$  0.032$\\

\hline
\end{tabular}
\end{table*}
\begin{table*}
\addtocounter{table}{-1}
\caption{\it Continued.}
\centering
\begin{tabular}{@{}l  r  r r r r r r r r r@{}}
\hline\hline
Galaxy       & $\log$(age)&   [M/H]&  $\log$(age)&[M/H]&$\log$(age)&[M/H]&$\log$(age)&[M/H]&$\log$(age)&   [M/H]\\
              &\multicolumn{2}{c}{[MgbFe]--H$\beta$}&
              \multicolumn{2}{c}{Fe4383--H$\beta$}&\multicolumn{2}{c}{Mgb--H$\beta$}&\multicolumn{2}{c}{CN$_2$--H$\beta$}&
              \multicolumn{2}{c}{(spectral synthesis)}\\
\hline
NGC 3665      &10.105& 0.014  &  10.159&$  -0.131$&    9.967&   0.251&    10.003&     0.181& 9.910&$  0.045$\\
              & 0.132& 0.188  &   0.087&$   0.100$&    0.152&   0.153&     0.108&     0.091& 0.065&$  0.052$\\
NGC 3818      & 9.761& 0.218  &   9.870&$   0.018$&    9.729&   0.310&     9.597&     0.444&10.077&$  0.099$\\
              & 0.090& 0.071  &   0.055&$   0.059$&    0.147&   0.117&     0.083&     0.065& 0.039&$  0.033$\\
NGC 4261      & 9.709& 0.423  &   9.879&$   0.075$&         &        &     9.582&     0.544&10.095&$  0.118$\\
              & 0.141& 0.069  &   0.047&$   0.053$&         &        &     0.140&     0.316& 0.034&$  0.032$\\
NGC 4278      &10.095& 0.152  &        &        &    9.931&   0.500&     9.968&       0.372&10.066&$  0.079$\\
              & 0.080& 0.072  &        &        &    0.063&   0.042&     0.038&       0.050& 0.039&$  0.031$\\
NGC 4365      & 9.900& 0.344  &   9.978&$   0.040$&         &        &     9.881&     0.414&10.100&$  0.139$\\
              & 0.084& 0.065  &   0.074&$   0.064$&         &        &     0.065&     0.139& 0.035&$  0.030$ \\
NGC 4374      &10.054& 0.118  &  10.148&$  -0.069$&    9.925&$   0.429$&     9.941&   0.354&10.017&$  0.099$\\
              & 0.103& 0.105  &   0.045&$   0.052$&    0.029&$   0.062$&     0.083&   0.075& 0.039&$  0.032$ \\
NGC 4415      & 9.855&-0.146  &   9.893&$  -0.256$&    9.841&$  -0.103$&          &        & 9.751&$ -0.424$\\
              & 0.104& 0.118  &   0.092&$   0.108$&    0.126&$   0.197$&          &        & 0.062&$  0.079$ \\
NGC 4431      &10.001&-0.408  &   9.972&$  -0.264$&    9.973&$  -0.270$&          &        & 9.955&$ -0.483$\\
              & 0.125& 0.243  &   0.179&$   0.174$&    0.146&$   0.325$&          &        & 0.067&$  0.069$ \\
NGC 4464      &10.057&-0.060  &  10.096&$  -0.154$&    9.943&$   0.164$&     9.929&   0.223& 9.846&$  0.115$\\
              & 0.085& 0.100  &   0.058&$   0.054$&    0.113&$   0.147$&     0.038&   0.056& 0.046&$  0.032$\\
NGC 4467      & 9.993& 0.027  &  10.043&$  -0.063$&    9.934&$   0.171$&     9.917&   0.254& 9.851&$  0.116$\\
              & 0.130& 0.135  &   0.122&$   0.113$&    0.156&$   0.180$&     0.119&   0.134& 0.046&$  0.031$\\
NGC 4472      & 9.984& 0.220  &  10.072&$   0.068$&    9.935&$   0.355$&     9.921&   0.429&10.098&$  0.139$\\
              & 0.084& 0.078  &   0.098&$   0.094$&    0.056&$   0.090$&     0.077&   0.050& 0.035&$  0.030$\\
NGC 4478      & 9.874& 0.055  &   9.885&$  -0.020$&    9.878&$   0.029$&     9.851&   0.199& 9.699&$  0.093$\\
              & 0.067& 0.088  &   0.049&$   0.051$&    0.109&$   0.168$&     0.081&   0.111& 0.044&$  0.034$\\
NGC 4486b     &      &        &        &$        $&    9.981&   0.403&     9.993&   0.380& 9.989&$  0.130$\\
              &      &        &        &$        $&    0.096&   0.099&     0.069&   0.064& 0.044&$  0.031$\\
NGC 4489      &9.437 & 0.223  &   9.429&$   0.263$&    9.206&   0.426&     9.429&   0.265& 9.787&$  0.149$\\
              &0.126 & 0.179  &   0.123&$   0.156$&    0.276&   0.149&     0.270&   0.230& 0.045&$  0.028$\\
NGC 4552      &10.094& 0.100  &        &$        $&    9.933&   0.431&     9.752&   0.488& 9.949&$  0.121$\\
              & 0.091& 0.115  &        &$        $&    0.034&   0.064&     0.258&   0.097& 0.045&$  0.031$\\
NGC 4564      & 9.904& 0.256  &   9.931&$   0.109$&    9.854&   0.459&          &        &10.025&$  0.117$\\
              & 0.031& 0.055  &   0.042&$   0.046$&    0.100&   0.049&          &        & 0.046&$  0.032$\\
NGC 4594      &      &        &  10.102&$  -0.066$&    9.923&   0.334&     9.805&   0.424&10.101&$  0.118$\\
              &      &        &   0.152&$   0.212$&    0.169&   0.151&     0.203&   0.279& 0.033&$  0.031$\\
NGC 4621      &9.942 & 0.299  &  10.140&$  -0.115$&    9.917&   0.426&     9.772&   0.442&10.057&$  0.115$\\
              &0.038 & 0.050  &   0.044&$   0.048$&    0.027&   0.063&     0.137&   0.053& 0.044&$  0.032$\\
NGC 4636      &10.011& 0.181  &  10.158&$  -0.104$&    9.918&   0.458&     9.770&   0.463& 9.894&$  0.122$\\
              &0.121 & 0.104  &   0.090&$   0.105$&    0.042&   0.073&     0.220&   0.081& 0.047&$  0.031$ \\

\hline
\end{tabular}
\end{table*}

\begin{table*}
\addtocounter{table}{-1}
\caption{\it Continued.}
\centering
\begin{tabular}{@{}l  r  r r r r r r r r r@{}}
\hline\hline
Galaxy       & $\log$(age)&   [M/H]&  $\log$(age)&[M/H]&$\log$(age)&[M/H]&$\log$(age)&[M/H]&$\log$(age)&[M/H]\\
              &\multicolumn{2}{c}{[MgbFe]--H$\beta$}&
              \multicolumn{2}{c}{Fe4383--H$\beta$}&\multicolumn{2}{c}{Mgb--H$\beta$}&\multicolumn{2}{c}{CN$_2$--H$\beta$}&
               \multicolumn{2}{c}{(spectral synthesis)}\\
\hline
NGC 4673      &9.577 & 0.297  &   9.722&$   0.090$&    9.556&   0.434&     9.552&   0.465& 9.891&$  0.059$\\
              &0.207 & 0.164  &   0.085&$   0.071$&    0.027&   0.061&     0.034&   0.197& 0.066&$  0.051$\\
NGC 4692      &9.908 & 0.305  &   9.948&$   0.112$&    9.872&   0.456&     9.726&   0.457& 9.984&$  0.131$\\
              &0.127 & 0.083  &   0.121&$   0.106$&    0.052&   0.077&     0.196&   0.111& 0.045&$  0.031$\\
NGC 4697      &9.770 & 0.197  &   9.833&$   0.138$&    9.780&   0.186&     9.599&   0.428&10.042&$  0.108$\\
              &0.082 & 0.072  &   0.065&$   0.045$&    0.092&   0.131&     0.082&   0.064& 0.044&$  0.033$\\
NGC 4742      &9.018 & 0.489  &   9.115&$   0.368$&         &        &          &        & 9.509&$ -0.007$\\
              &0.109 & 0.093  &   0.003&$   0.024$&         &        &          &        & 0.064&$  0.075$\\
NGC 4839      &9.732 & 0.423  &   9.915&$   0.241$&         &        &     9.732&   0.425& 9.911&$  0.126$\\
              &0.205 & 0.097  &   0.178&$   0.141$&         &        &     0.235&   0.132& 0.041&$  0.031$\\
NGC 4842A     &9.925 & 0.327  &  10.087&$  -0.024$&    9.904&   0.460&     9.945&   0.239& 9.953&$  0.120$\\
              &0.107 & 0.079  &   0.116&$   0.122$&    0.033&   0.063&     0.154&   0.162& 0.045&$  0.031$\\
NGC 4842B     &9.934 & 0.189  &   9.984&$   0.060$&    9.659&   0.402&     9.947&   0.139& 9.783&$  0.113$\\
              &0.140 & 0.121  &   0.143&$   0.134$&    0.251&   0.112&     0.145&   0.129& 0.046&$  0.032$\\
NGC 4875      &9.922 & 0.182  &  10.050&$  -0.123$&    9.732&   0.419&     9.920&   0.196& 9.973&$  0.122$\\
              &0.103 & 0.101  &   0.129&$   0.106$&    0.191&   0.112&     0.150&   0.174& 0.051&$  0.031$\\
NGC 4864      &9.885 & 0.272  &   9.952&$  -0.029$&    9.867&   0.348&     9.892&   0.236& 9.852&$  0.077$\\
              &0.117 & 0.101  &   0.171&$   0.150$&    0.151&   0.144&     0.156&   0.144& 0.072&$  0.052$\\
NGC 4865      &9.726 & 0.340  &   9.888&$  -0.023$&    9.586&   0.513&     9.723&   0.353& 9.830&$  0.128$\\
              &0.159 & 0.077  &   0.104&$   0.096$&    0.375&   0.098&     0.172&   0.113& 0.047&$  0.031$\\
NGC 4867      &9.720 & 0.294  &   9.845&$  -0.039$&    9.572&   0.517&     9.700&   0.387& 9.805&$  0.012$\\
              &0.256 & 0.216  &   0.142&$   0.143$&    0.178&   0.096&     0.177&   0.140& 0.075&$  0.064$\\
NGC 4874      &9.923 & 0.253  &   9.963&$   0.109$&    9.893&   0.416&     9.818&   0.411& 9.883&$  0.133$\\
              &0.088 & 0.091  &   0.140&$   0.131$&    0.053&   0.085&     0.148&   0.182& 0.045&$  0.030$\\
NGC 4889      &9.963 & 0.302  &  10.190&$  -0.121$&    9.912&   0.523&     9.752&   0.474& 9.988&$  0.131$\\ 
              &0.187 & 0.114  &   0.166&$   0.231$&    0.047&   0.062&     0.297&   0.108& 0.044&$  0.031$\\
NGC 4908      &9.977 & 0.077  &  10.086&$  -0.135$&    9.910&   0.323&     9.954&   0.123& 9.923&$  0.091$\\
              &0.127 & 0.123  &   0.121&$   0.099$&    0.119&   0.125&     0.064&   0.120& 0.041&$  0.032$\\
NGC 5638      &9.715 & 0.232  &   9.744&$   0.104$&    9.591&   0.337&     9.565&   0.477& 9.843&$  0.147$\\
              &0.073 & 0.069  &   0.051&$   0.058$&    0.100&   0.100&     0.030&   0.297& 0.045&$  0.029$\\
NGC 5796      &9.852 & 0.235  &   9.872&$   0.120$&    9.582&   0.542&     9.588&   0.515& 9.942&$  0.145$\\
              &0.103 & 0.150  &   0.089&$   0.082$&    0.046&   0.118&     0.148&   0.308& 0.045&$  0.029$\\
NGC 5812      &9.728 & 0.310  &   9.843&$   0.144$&    9.705&   0.413&          &        &10.037&$  0.128$\\
              &0.089 & 0.057  &   0.068&$   0.050$&    0.071&   0.084&          &        & 0.045&$  0.032$\\
NGC 5813      &9.741 & 0.242  &   9.855&$   0.036$&    9.704&   0.403&     9.575&   0.537&10.084&$  0.110$\\
              &0.087 & 0.064  &   0.056&$   0.059$&    0.072&   0.086&     0.026&   0.192&0.039 &$  0.032$ \\
NGC 5831      &9.543 & 0.316  &   9.556&$   0.226$&    9.541&   0.331&     9.511&   0.516&9.937 &$  0.127$\\
              &0.130 & 0.160  &   0.032&$   0.048$&    0.093&   0.164&     0.041&   0.203& 0.046&$  0.031$\\

\hline
\end{tabular}
\end{table*}
\begin{table*}
\addtocounter{table}{-1}
\caption{\it Continued.}
\centering
\begin{tabular}{@{}l  r  r r r r r r r r r@{}}
\hline\hline
Galaxy       & $\log$(age)&   [M/H]&  $\log$(age)&[M/H]&$\log$(age)&[M/H]&$\log$(age)&[M/H]&$\log$(age)&[M/H]\\
              &\multicolumn{2}{c}{[MgbFe]--H$\beta$}&
              \multicolumn{2}{c}{Fe4383--H$\beta$}&\multicolumn{2}{c}{Mgb--H$\beta$}&\multicolumn{2}{c}{CN$_2$--H$\beta$}&
               \multicolumn{2}{c}{(spectral synthesis)}\\
\hline
NGC 5845      &9.919 & 0.172  &   9.984&$  -0.018$&    9.917&   0.183&     9.723&   0.445&10.065&$  0.112$\\
              &0.064 & 0.108  &   0.068&$   0.056$&    0.081&   0.139&     0.131&   0.092& 0.044&$  0.033$\\
NGC 5846      &9.926 & 0.219  &  10.029&$  -0.025$&    9.737&   0.421&     9.884&   0.429&10.049&$  0.103$\\
              &0.048 & 0.061  &   0.089&$   0.079$&    0.134&   0.087&     0.070&   0.142& 0.044&$  0.032$\\
NGC 5846A     &10.029& 0.077  &  10.162&$  -0.230$&    9.925&   0.338&     9.931&   0.311& 9.893&$  0.122$\\
              & 0.103& 0.111  &   0.064&$   0.076$&    0.052&   0.092&     0.074&   0.076& 0.047&$  0.031$\\
NGC 6127      & 9.927& 0.273  &  10.004&$   0.072$&    9.881&   0.496&     9.734&   0.453& 9.925&$  0.136$\\
              & 0.042& 0.061  &   0.095&$   0.083$&    0.039&   0.079&     0.190&   0.078& 0.047&$  0.030$\\
NGC 6166      &      &        &        &$        $&    9.989&$   0.536$&    10.197&   0.310&10.099&$  0.001$\\
              &      &        &        &$        $&    0.352&$   0.097$&     0.197&   0.166& 0.038&$  0.047$\\
NGC 6411      &9.718 & 0.252  &   9.723&$   0.228$&    9.589&$   0.378$&     9.576&   0.447&10.104&$  0.107$\\
              &0.093 & 0.078  &   0.098&$   0.071$&    0.104&$   0.097$&     0.094&   0.090& 0.035&$  0.033$\\
NGC 6482      &      &        &        &$        $&   10.040&$   0.445$&    10.162&   0.364&10.058&$  0.116$\\
              &      &        &        &$        $&    0.130&$   0.025$&     0.076&   0.023& 0.044&$  0.032$\\
NGC 6577      &9.947 & 0.279  &  10.137&$  -0.107$&    9.913&$   0.448$&     9.936&   0.326& 9.953&$  0.116$\\
              &0.073 & 0.071  &   0.097&$   0.103$&    0.045&$   0.077$&     0.089&   0.085& 0.045&$  0.031$\\
NGC 6702      &      &        &   9.211&$   0.451$&         &        &       9.314&   0.538& 9.996&$  0.119$\\
              &      &        &   0.233&$   0.067$&         &        &       0.130&   0.094& 0.046&$  0.031$\\
NGC 6703      & 9.745& 0.092  &   9.721&$   0.197$&    9.776&$   0.012$&     9.570&   0.442& 9.945&$  0.125$\\
              & 0.097& 0.092  &   0.069&$   0.047$&    0.115&$   0.168$&     0.027&   0.087& 0.045&$  0.031$\\
NGC 7052      &      &        &        &$        $&         &        &            &        &10.073&$  0.096$\\
              &      &        &        &$        $&         &        &            &        & 0.038&$  0.032$\\
IC 767        & 9.397&-0.478  &   9.344&$  -0.107$&    9.395&$  -0.468$&          &        & 9.279&$ -0.240$\\
              & 0.173& 0.296  &   0.131&$   0.308$&    0.245&$   0.630$&          &        & 0.045&$  0.102$\\
IC 794        & 9.731&-0.184  &   9.581&$   0.083$&    9.774&$  -0.413$&     9.807&   0.442& 9.788&$ -0.072$\\
              & 0.175& 0.188  &   0.176&$   0.156$&    0.501&$   0.366$&     0.120&   0.042& 0.069&$  0.059$\\
IC 832        & 9.797& 0.091  &   9.889&$  -0.161$&    9.720&$   0.290$&     9.827&   0.026& 9.756&$ -0.066$\\
              & 0.097& 0.109  &   0.103&$   0.099$&    0.158&$   0.138$&     0.266&   0.119& 0.068&$  0.065$\\
IC 3957       & 9.957& 0.108  &  10.009&$   0.011$&    9.743&$   0.400$&     9.915&   0.280&10.150&$  0.162$\\
              & 0.126& 0.116  &   0.115&$   0.113$&    0.147&$   0.099$&     0.111&   0.138& 0.029&$  0.026$\\
IC 3959       & 9.733& 0.324  &   9.895&$  -0.003$&         &        &       9.703&   0.455&10.083&$  0.112$\\
              & 0.111& 0.063  &   0.081&$   0.081$&         &        &       0.168&   0.114& 0.039&$  0.032$\\
IC 3963       & 9.721& 0.211  &   9.737&$   0.141$&    9.591&$   0.347$&     9.727&   0.187& 9.837&$  0.021$\\
              & 0.144& 0.106  &   0.143&$   0.115$&    0.152&$   0.124$&     0.150&   0.149& 0.077&$  0.062$\\
IC 3973       & 9.569& 0.342  &   9.595&$   0.211$&    9.546&$   0.502$&     9.547&   0.500& 9.834&$  0.133$\\
              & 0.103& 0.061  &   0.111&$   0.085$&    0.033&$   0.048$&     0.041&   0.204& 0.046&$  0.030$\\
\hline
\end{tabular}
\end{table*}
\begin{table*}
\addtocounter{table}{-1}
\caption{\it Continued.}
\centering
\begin{tabular}{@{}l  r  r r r r r r r r r@{}}
\hline\hline
Galaxy       & $\log$(age)&   [M/H]&  $\log$(age)&[M/H]&$\log$(age)&[M/H]&$\log$(age)&[M/H]&$\log$(age)&[M/H]\\
              &\multicolumn{2}{c}{[MgbFe]--H$\beta$}&
              \multicolumn{2}{c}{Fe4383--H$\beta$}&\multicolumn{2}{c}{Mgb--H$\beta$}&\multicolumn{2}{c}{CN$_2$--H$\beta$}&
               \multicolumn{2}{c}{(spectral synthesis)}\\
\hline
IC 4026       & 9.723&$ 0.270$ &   9.825&$   0.021$&    9.707&$   0.347$&      9.773&   0.128&10.149&$  0.149$\\
              & 0.259&$ 0.223$ &   0.147&$   0.156$&    0.181&$   0.140$&      0.158&   0.206& 0.029&$  0.028$\\
IC 4042       &      &$      $  &        &       &      10.216&$   0.188$&          &        &10.151&$  0.152$\\
              &      &$      $  &        &       &       0.149&$   0.141$&          &        & 0.029&$  0.028$\\
IC 4051       &9.965 &$ 0.315$  &        &        &      9.912&$   0.540$&     9.921&   0.489& 9.911&$  0.130$ \\
              &0.152 &$ 0.099$  &        &        &      0.034&$   0.056$&     0.038&   0.043& 0.041&$  0.030$\\
CGCG 159-41   &10.112&$ 0.089$  &        &        &      9.936&$   0.435$&     9.970&   0.327& 9.890&$  0.127$ \\
              &0.113 &$ 0.141$  &        &        &      0.047&$   0.073$&     0.130&   0.115& 0.047&$  0.031$\\
CGCG 159-43   &9.994 &$ 0.188$  &  10.144&$  -0.112$&    9.915&$   0.451$&     9.919&   0.427&10.148&$  0.155$ \\
              &0.106 &$ 0.095$  &   0.074&$   0.085$&    0.037&$   0.071$&     0.086&   0.062& 0.028&$  0.027$\\
CGCG 159-83   &      &$      $  &        &        &         &        &              &        & 9.993&$ -0.172$ \\
              &      &$      $  &        &        &         &        &              &        & 0.067&$  0.068$\\
CGCG 159-89   & 9.573&$ 0.258$  &   9.704&$   0.122$&    9.558&$   0.352$&     9.559&   0.343&10.101&$  0.114$ \\
              & 0.218&$ 0.189$  &   0.105&$   0.108$&    0.116&$   0.097$&     0.215&   0.174& 0.034&$  0.032$\\
DRCG 27-032   &10.187&$-0.137$  &        &        &     10.177&$  -0.119$&          &        &10.014&$ -0.143$ \\
              &0.219 &$ 0.415$  &        &        &      0.226&$   0.650$&          &        & 0.061&$  0.067$\\
DRCG 27-127   &9.703 &$ 0.323$  &   9.784&$   0.052$&    9.572&$   0.471$&     9.762&   0.097&10.106&$  0.108$ \\
              &0.172 &$ 0.104$  &   0.190&$   0.171$&    0.219&$   0.096$&     0.160&   0.169& 0.034&$  0.032$\\
DRCG 27-128   &9.585 &$ 0.157$  &   9.580&$   0.181$&    9.556&$   0.320$&          &        & 9.960&$  0.092$ \\
              &0.154 &$ 0.121$  &   0.146&$   0.119$&    0.159&$   0.155$&          &        & 0.045&$  0.032$\\
GMP 3121      &10.104&$-0.247$  &   9.964&$   0.064$&   10.115&$  -0.291$&          &        & 9.610&$ -0.050$ \\
              &0.222 &$ 0.381$  &   0.193&$   0.307$&    0.248&$   0.695$&          &        & 0.071&$  0.082$\\
GMP 3196      &9.936 &$-0.339$  &   9.856&$  -0.088$&    9.874&$  -0.138$&          &        & 9.655&$ -0.325$ \\
              &0.284 &$ 0.229$  &   0.195&$   0.180$&    0.229&$   0.286$&          &        & 0.072&$  0.094$\\
MCG+05-31-63  &10.053&$-0.061$  &  10.130&$  -0.291$&    9.933&$   0.196$&          &        & 9.920&$ -0.107$ \\
              &0.142 &$ 0.214$  &   0.149&$   0.201$&    0.229&$   0.208$&          &        & 0.073&$  0.071$\\
PGC 126756    &9.957 &$-0.507$  &   9.905&$  -0.314$&    9.969&$  -0.552$&          &        & 9.656&$ -0.329$ \\
              &0.176 &$ 0.363$  &   0.297&$   0.256$&    0.190&$   0.652$&          &        & 0.072&$  0.093$\\
PGC 126775    &9.411 &$-0.275$  &   9.380&$  -0.061$&    9.396&$  -0.163$&     9.307&   0.311& 9.559&$ -0.369$ \\
              &0.249 &$ 0.387$  &   0.222&$   0.359$&    0.359&$   0.802$&     0.281&   0.174& 0.059&$  0.099$\\
RB 91         &9.873 &$ 0.127$  &   9.880&$   0.084$&    9.726&$   0.353$&     9.880&   0.086&10.151&$  0.152$ \\
              &0.190 &$ 0.154$  &   0.201&$   0.172$&    0.212&$   0.152$&     0.145&   0.254& 0.029&$  0.028$\\
RB 113        &      &         &        &        &         &        &              &        &10.153&$  0.156$ \\
              &      &         &        &        &         &        &              &        & 0.029&$  0.027$ \\
\hline
\end{tabular}
\end{table*}

\section{The SSP-parameters $\sigma$ relation}
 \label{vic-global}
\begin{figure*}
  \resizebox{0.6\hsize}{!}{\includegraphics[angle=0]{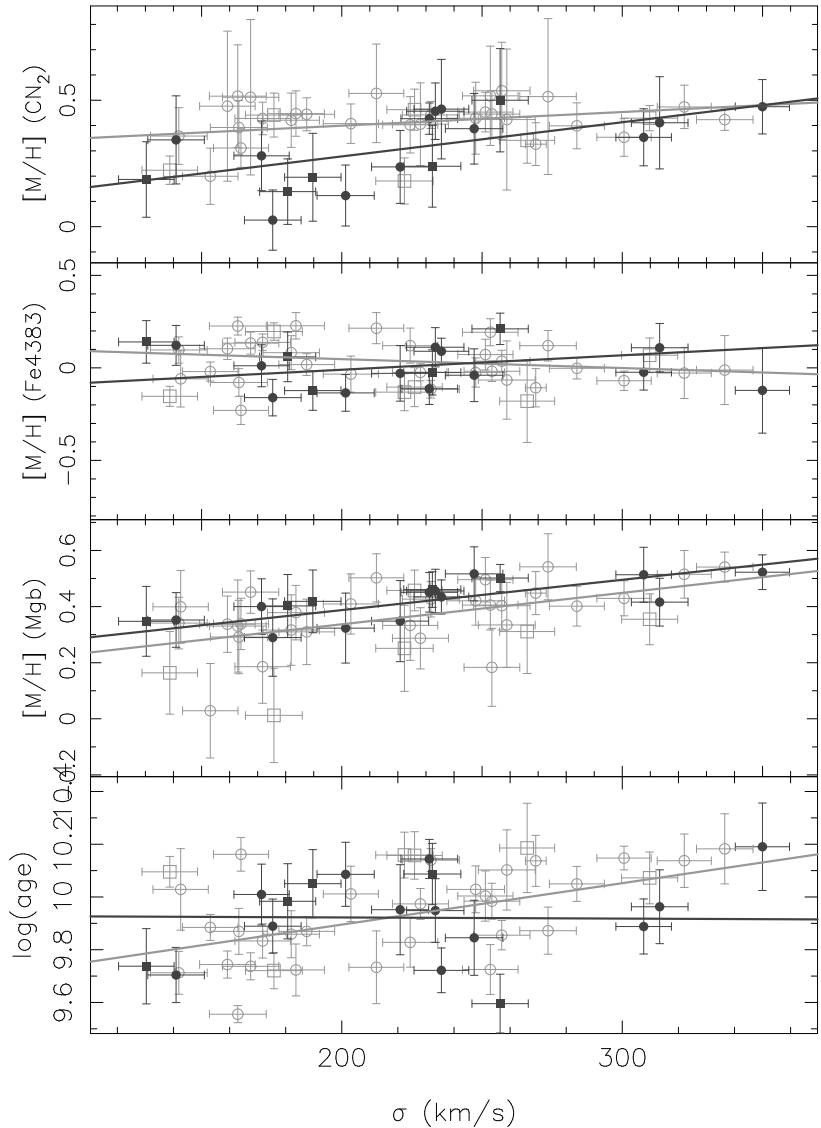}}
  \caption{Relations between metallicities, obtained with different indicators,
  and age against velocity dispersion for the sample of galaxies. Open symbols
  represent galaxies in low-density environments (LDEGs), while filled symbols
  indicate galaxies in high density environments (HDEGs).  Squares correspond
  to S0 galaxies while elliptical galaxies are represented with circles. Grey
  and black lines show the linear fit, weighting with the errors in both axes,
  to the LDEGs and HDEGs respectively. \label{panel}}
\end{figure*}

Fig. \ref{panel} shows the ages and metallicities obtained in different
index--index diagrams as a function of the velocity dispersion for both HDEGs
(black symbols) and LDEGs (grey symbols).  We have not plotted the dwarfs
galaxies of the sample, since it is not clear wheter these galaxies are the
faint extension of the giant ellipticals (see, for example Gorgas et al.\ 1997;
Pedraz et al.\ 2002; Graham \& Guzm\'an 2003). Error-weighted linear fits to
each subsample are also shown in the figure. 
 These fits were dervied by initially performing an  unweighted 
ordinary least-squares
regression of Y on X  and the 
coefficients from the first fit were then employed to derive (numerically, with
a downhill method) the straight line data-fit with errors in both coordinates.
Table \ref{mass-metallicity-fits}
lists the slopes of the fits with their corresponding errors. The last two
columns indicate the $t$--statistic obtained in the comparison of the slopes of
HDEGs and LDEGs, and the probability that the slopes are different by chance.
We  analyse each panel separately:

\begin{itemize}
  \item $[$M/H$]$ (CN$_2$--H$\beta$) versus $\sigma$: While there is a
  correlation for the HDEGs between the metallicity and the velocity
  dispersion, 
  this is not true for the LDEGs, which are compatible with a null relation. 
  The most massive galaxies ($\sigma > 300$ km s$^{-1}$), however, show the same
  behaviour in both environments, while  for the rest of the galaxies, there
  exists a clear difference between the  metallicity in the two galaxy
  subsamples.

  \item $[$M/H$]$ (Fe4383--H$\beta$) versus $\sigma$: This panel shows a
  similar behaviour to the former one, being the correlation between the
  metallicity measured with Fe4383 and the velocity dispersion stronger for
  HDEGs.  For the LDEGs, however, the relation is almost flat and even
  slightly negative. 

  \item $[$M/H$]$ (Mgb-H$\beta$) versus $\sigma$: The relation between the
  metallicity measured with Mgb and the velocity dispersion is rather similar
  for both subsamples of galaxies. Furthermore, we did not find any significant
  difference between the zero point of both relations; both samples are
  compatible with the same metallicity (as measured with Mgb) $\sigma$
  relation.

  \item Age versus $\sigma$: In this panel it can be seen that, while for the
  HDEGs the relation is flat, there exists a correlation between the age and
  velocity dispersion for the LDEGs, in the sense that low velocity dispersion
  galaxies tend to be younger. This is in agreement with the suggestion by
  Trager et al.\ (2000b), who found differences in the ($\sigma$, t) plane
  between galaxies in the field and in the Fornax cluster.  Interestingly, J\o
  rgensen (1999) did not find any correlation between age and velocity
  dispersion in her study of a sample of galaxies in the Coma cluster, although
  she found a considerable dispersion in the ages of the galaxies.  Caldwell 
  et al.\ (2003) also found younger ages for lower $\sigma$ galaxies in a sample
  of Virgo galaxies and galaxies in lower environments. Thomas et
  al.\ (2004) did not find a significant trend between the age and the velocity
  dispersion in either their sample of high- or low-density environment
  galaxies. However, they argued that the correlated errors of age and
  metallicity tend to dilute a correlation between age and the velocity
  dispersion and that their observational  data are best reproduced by a
  relatively flat, but significant correlation.

  It is also clear from Fig. \ref{panel}  that the age dispersion for less
  massive galaxies is higher than for the more massive ones, in agreement with
  other studies (e.g.  Bender, Burstein \& Faber 1993; Bressan, Chiosi \&
  Tantalo 1996; Worthey 1996; Mehlert et al.\ 1998; Vazdekis \& Arimoto 1999;
  J\o rgensen 1999; Smail et al.\ 2001, Caldwell et al.\ 2003).
\end{itemize}

\begin{table*}
  \caption{Slopes of the linear fits ($b$) and their errors ($\Delta b$) of the
  different parameters in Fig. \ref{panel} with the central velocity
  dispersion.  The left side of the table show the values for the LDEGs, while
  the right side of the table, the values for the HDEGs. The last two columns
  show the result of a $t$-test to verify if the slopes of both subsamples of
  galaxies are different. In particular, the last column shows the probability
  that the slopes of HDEGs and LDEGs are different by chance.
  \label{mass-metallicity-fits}}
 \centering
 \begin{tabular}{@{}l r r| r r r r@{}}
  \hline\hline
                  & \multicolumn{2}{c}{LDEG} & \multicolumn{2}{c}{HDEG}& &\\
  \hline
                  &\multicolumn{1}{c}{$b$}&\multicolumn{1}{c}{$\Delta
  b$}&\multicolumn{1}{c}{$b$}&\multicolumn{1}{c}{$\Delta b$} &
  \multicolumn{1}{c}{$t$}&\multicolumn{1}{c@{}}{$\alpha$}\\  
 $[M/H$] (CN$_2$) &$  0.00037$ & 0.00034       & 0.00138  & 0.00065    & 1.37&9\%\\
 $[M/H]$ (Fe4383) &$ -0.00050$ & 0.00030       & 0.00085  & 0.00053    & 2.21&1\%\\
 $[M/H]$ (Mgb)    &$  0.00113$ & 0.00031       & 0.00077  & 0.00042    & 0.69&24\%\\
 log (age)        &$  0.00177$ & 0.00033       & 0.00004  & 0.00064    & 2.41&1\%\\
  \hline
  \end{tabular}
\end{table*}

\begin{figure}
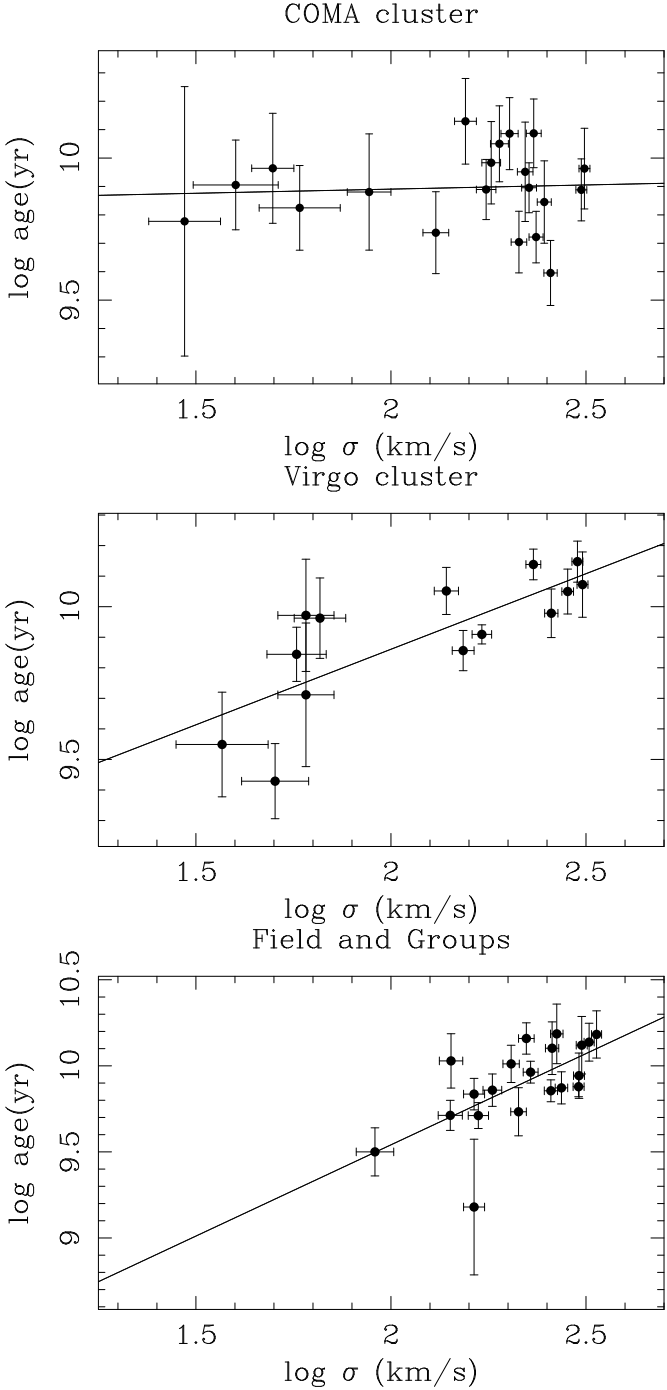

\resizebox{1.0\hsize}{!}{\includegraphics[angle=-90]{coma.ages.ps}}
\resizebox{1.0\hsize}{!}{\includegraphics[angle=-90]{virgo.ages.ps}}
\resizebox{1.0\hsize}{!}{\includegraphics[angle=-90]{field.ages.ps}}
\caption{Relation of the age with the velocity dispersion for galaxies in 
the Coma cluster, Virgo cluster, and galaxies in the field and poor groups.\label{ages.env}}
\end{figure}

 As stated in Paper I, a decision was made to include the Virgo galaxies 
with the 
rest of the LDEGs in order to enlarge the the number of galaxies 
in this group, as we did not find any
difference between the stellar population of those two sub-samples. 
In  further support of this decision,  we have plotted, 
in Fig. \ref{ages.env}, the 
relation of the ages with 
the velocity dispersion for galaxies in the Coma cluster, Virgo clusters 
and groups 
separately. In opposition with the behaviour of galaxies in the 
Fornax  (Kuntschner 2000) and Coma cluster, galaxies in the Virgo 
cluster show a greater spread in the ages of their populations and 
they appear to be correlated
with the velocity dispersion of the galaxies.

\section{Simple stellar population or multiple bursts?}
\label{ss.sb}
\begin{figure}
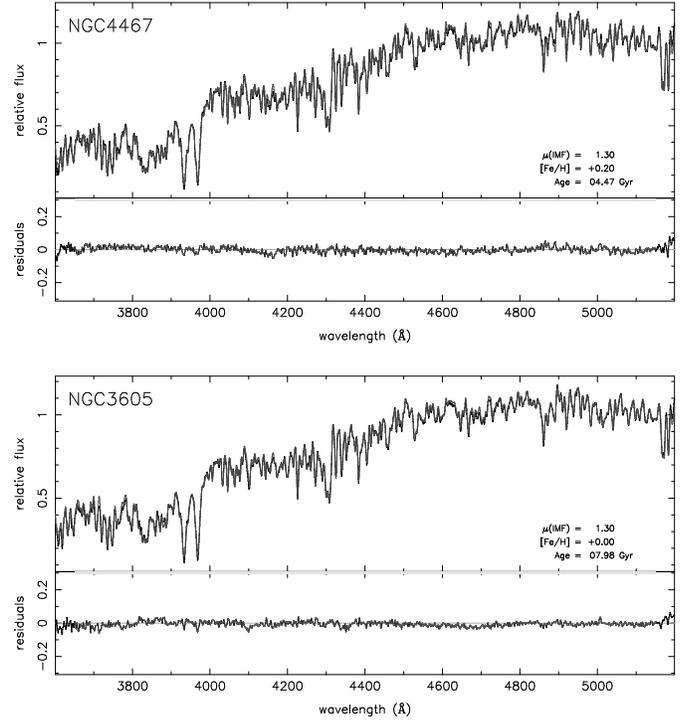
 
\centering
\resizebox{1.0\hsize}{!}{\includegraphics[angle=-90]{n4467.apeq_3650_5150_sp.ps}}

  \vspace{4mm}

 \resizebox{1.0\hsize}{!}{\includegraphics[angle=-90]{n3605.apeq_3650_5150_sp.ps}}
  \caption{Comparison of the observed spectra of the galaxies NGC 4467  and NGC
  3605 (black line) and the synthetic spectra from V06 (grey line). The lower
  panels show the difference between both spectra. The parameters of the
  synthetic spectra are shown in the panel.\label{comparison.synthetic}}
\end{figure}

The new generation of stellar population models do not only predict individual
features as a function of the age and metallicity, but  synthesise full
spectral energy distributions (SEDs) (Vazdekis 1999; Vazdekis et al.\ 2003;
Bruzual \& Charlot 2003; Le Borgne et al.\ 2004).  This
is possible thanks to the growing number of stellar libraries with a large
number of stars (eg. LeBorgne et al. 2003; Valdes et al.\ 2004;
S\'anchez--Bl\'azquez et al.\ 2006). To show the potential of these new
models, in Fig. \ref{comparison.synthetic} we compare the spectral energy
distribution of two galaxies, NGC 4467 and NGC 3605, with two synthetic spectra
from V06 for a single population of age and metallicity as indicate in the
panels. Both observed and synthetic spectra were  normalised to 5000 \AA. As
can be seen, the coincidence is very good.  The shape of both spectra (observed
and synthetic) are real and their similarity gives support to the flux
calibration of both the galaxies and the stellar spectra. 

\begin{figure} 
\centering
  \resizebox{1.0\hsize}{!}{\includegraphics[angle=-90]{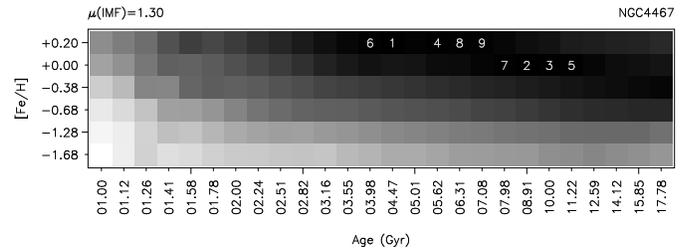}}
  \caption{Grey-scale diagram of the standard deviation in the residuals of the
  comparison of different synthetic spectra with the spectrum of NGC 4467.
  Darker squares represent lower residuals. The numbers indicate the
  combination of age and metallicity of the best 9 solutions, ordered from
  the lowest standard deviation (1) to the highest (9).\label{def-age-meta}}
\end{figure}

The age and metallicity values of the synthetic spectra that best reproduce the
observed spectra were calculated by minimising the residual r.m.s.\ (obtained
as the differences between the observed and the synthetic spectra in the region
3650--5150 \AA).  Internal reddening has not been taken into account
in this analysis.  
The estimation of the synthetic spectra that best reproduce
the observed  is affected by the age--metallicity degeneracy. To show
this, we have represented, in Fig. \ref{def-age-meta}, a grey-scale diagram 
with the r.m.s.\ of the residuals in the comparison of the synthetic spectra of
different ages and metallicities with the spectrum of the  galaxy NGC 4467. 
For this reason, 
we have computed the age and the metallicity by averaging the best 9 solutions
weighted with
the inverse of the residual variance of the comparison with the synthetic
spectra. The last column of Table \ref{edades.metalicidades} shows the ages
and metallicities obtained in this way for the sample of galaxies. 

\begin{figure}
\resizebox{1.0\hsize}{!}{\includegraphics[angle=-90]{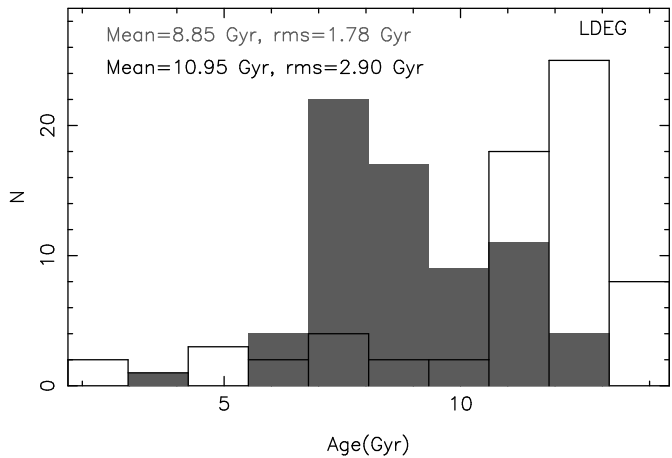}}
  \caption{Distribution of ages obtained comparing the synthetic spectra from
  V06 with the spectral energy distribution of LDEGs. The empty histogram show
  the ages obtained with the comparison in the spectral range
  \mbox{4750--5150 \AA}\ while the shaded histogram, the ages obtained
  comparing the region from \mbox{3650--4050 \AA}.\label{different}}
\end{figure}

\begin{figure}
\resizebox{1.0\hsize}{!}{\includegraphics[angle=-90]{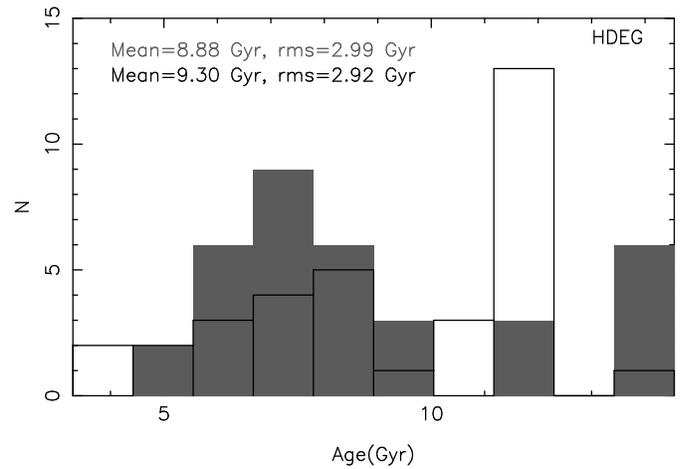}} 
  \caption{Distribution of ages obtained by comparing the synthetic spectra
  from the models of V06 with the observed spectra for the sample of HDEGs. The
  empty histogram represents the ages obtained in the spectral range
  \mbox{4750--5150 \AA}\ while the shaded histogram shows the calculated ages
  in the spectral range \mbox{3650--4050 \AA}.\label{different.hdeg}}
\end{figure}

It has been seen in Sec. \ref{vic-global} that
LDEGs span a broad range in their apparent mean ages and that, in some cases,
these ages are very low.  Since the models assume a
unique burst of star formation, these low values can indicate either that these
galaxies are genuinely young, that is, that they formed all their stars
recently, or that most of their stars were formed at early epochs, but they
have undergone later episodes of star formation involving a certain percentage
of the total mass of the galaxy (see Trager et al.\ 2000b).  In the latter case,
the apparent mean age would depend on the relative light contributions of the
different components to the considered spectral range.  
To distinguish between the two scenarios, we carried out a
comparison of the observed galaxy spectra with the synthetic spectra extracted
from V06 models in two different wavelength ranges: \mbox{3650--4050 \AA}\ and
\mbox{4750--5150 \AA}. 
The ages and metallicities were  calculated combining the values of the 9
synthetic spectra that best reproduced the observed spectra in each wavelength
interval, as explained above.  Figs. \ref{different} and \ref{different.hdeg}
show the age distributions obtained in both wavelength ranges for LDEG and HDEG
respectively. It is apparent from Fig. \ref{different} that, for LDEGs,  there
is a significant difference between the ages obtained in the two different
regions of the spectra. This is difficult to understand if all the stars were
formed in a single burst, and it suggests that many LDEGs are composite systems
consisting of an underlying old population plus, at least, a later star
formation burst.

\begin{figure}
\centering
\resizebox{1.0\hsize}{!}{\includegraphics[angle=-90]{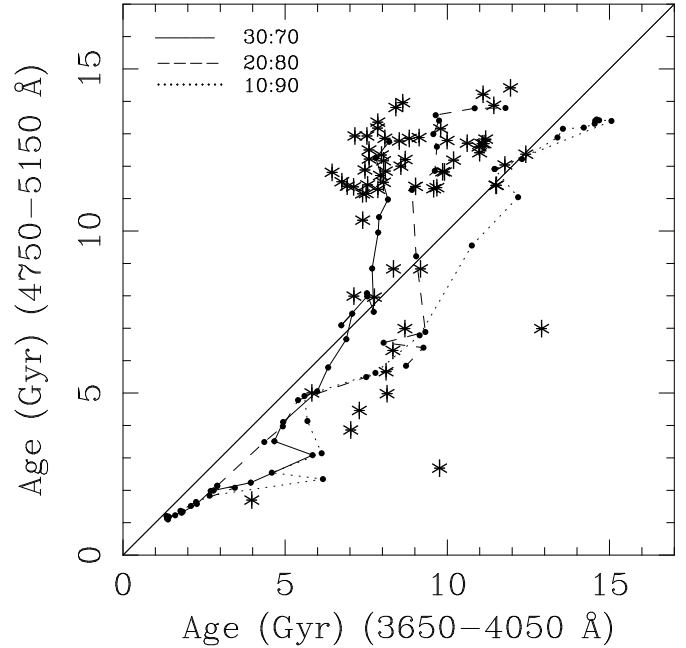}}
\caption{Comparison of the ages derived in two different spectral ranges
 using the models of V06. The asterisks are the values calculated for the
 LDEG spectra, while the filled circles are the ages derived from the  composite models in
 which two populations of different ages and metallicities are added (see text
 for details). The solid, dashed, and dotted lines connect the varios two-component
 model combinations for the 30:70, 20:80, and 10:90 young:old population mass ratios, 
 respectively. The age of the younger component increases from lower left to upper
 right of the diagram.\label{modelillo}}
\end{figure}

\begin{figure}
\resizebox{\hsize}{!}{\includegraphics[angle=-90]{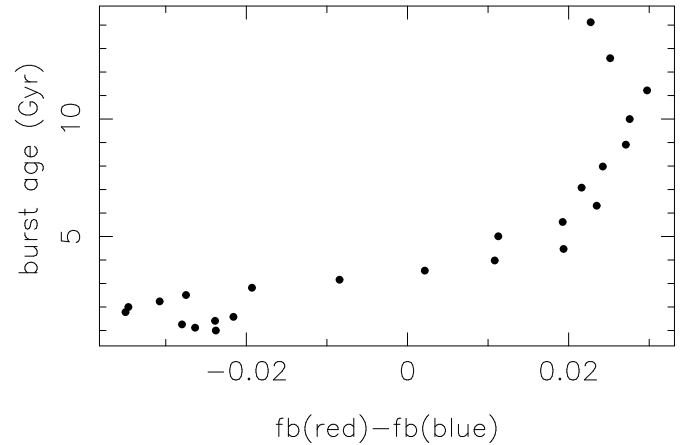}}
\caption{Differences  between the fraction of light contributed
by  a burst 
of star formation with  metallicity Z=$+$0.2 in two different 
spectral ranges (4750-51500\AA) and (3650-4050\AA) when 
added to a population of age=15.85 Gyr and metallicity Z$=-0.38$
for  different ages of the burst.\label{age.burst}}
\end{figure}

To study this in more detail, we have built different composite spectra in
which we have added to  an  old population of 15.85 Gyr and metallicity
$[$M/H$]$=$-0.38$ dex, different components of metallicity $[$M/H$]$=+0.2 and
ages ranging from 2.51 to 14.12 Gyr. The percentages of these two components
were  chosen to be  70 and 30\% (model 1, solid line),  
80 and 20\% (model 2, dashed line), and 90 and
10\% (model 3, dotted line) in  mass, respectively.   
Fig. \ref{modelillo} shows
the relation between the derived ages in these two spectral ranges for
different  models and,  over-plotted,  the derived ages for the LDEGs
(crosses). As can be seen, although is difficult to match the observed points
with single scenarios, as the contribution in mass and the look-back time of
the star formation event are highly degenerate, the combination of an old
population and a burst of star formation  would
lead to  similar trends in the derived ages as the observed for the LDEGs.  
We also note that, 
in order to reproduce the observed trends, 
the metallicity of
the young component must be higher than the metallicity of the underlying old
population, in agreement with the findings of other authors (eg. Ferreras,
Charlot \& Silk 1999; Trager et al.\ 2000a; Thomas et al.\ 2004).
 We note here  that the differences between the ages derived in the two 
spectral ranges do not follow simple relation. The shape of this relation 
depends on the difference in the light contribution of the burst to
the considered wavelength regions. Table \ref{tabl.fractions} shows 
these fractions, as  calculated in model 2. It can be noted 
that the contribution of the young population is higher in the red
wavelength range than in the blue when its age become older than 3.5 Gyr.
This is due to the larger metallicity of the young component with respect
to the subyacent population. In a model where both, subyacent population 
and burst have the same metallicity, the contribution of the young 
component to the total light is always larger in the bluer wavelengths.
Figure \ref{age.burst} shows the relation of the differences in  
the light fraction of the burst between  the two considered spectral regions, 
and the age of the burst, also calculated for model 2.
 As can be seen,  the differences 
are larger for older bursts (although the trend is not monotonic). That is
the reason why the differences between the ages calculated in 
two different spectral ranges  increase  with the age of the burst. 
\begin{table}
\begin{tabular}{l c  c}
\hline\hline
Age burst(Gyr) &  fb (3650-4050 \AA) & fb (4750-5150 \AA)\\
\hline
1.00           &    0.956            & 0.932            \\
1.12           &    0.944            & 0.917            \\
1.26           &    0.932            & 0.904           \\
1.41           &    0.920            & 0.896           \\
1.58           &    0.900            & 0.879           \\
1.78           &    0.852            & 0.817           \\
2.00           &    0.811            & 0.766           \\
2.24           &    0.766            & 0.712           \\
2.51           &    0.712            & 0.685           \\
2.82           &    0.653            & 0.634           \\
3.16           &    0.596            & 0.588            \\
3.55           &    0.544            & 0.547         \\
3.98           &    0.482            & 0.493         \\
4.47           &    0.430            & 0.449         \\
5.01           &    0.398            & 0.409        \\
5.62           &    0.343            & 0.362       \\
6.31           &    0.284            & 0.307       \\
7.08           &    0.246            & 0.268  \\
7.98           &    0.193            & 0.217  \\
8.91           &    0.161            & 0.188  \\
10.00          &    0.135            & 0.162  \\
11.22          &    0.110            & 0.140  \\
12.59          &    0.084            & 0.109  \\
14.12          &    0.070            & 0.093  \\
\hline
\end{tabular}
\caption{ Fraction of light contributed by a burst of 
star formation with metallicity z=+0.2 and strength of 20\% in  mass,
in a galaxy with an age 15.85 Gyr and metallicity $z=-0.38$ for 
different ages of the burst.\label{tabl.fractions}}
\end{table}

On the other hand, the distribution of ages for HDEGs (Fig.
\ref{different.hdeg}) shows no such a clear  dichotomy 
(see mean values in the insets). This is compatible with the idea that these
galaxies  constitute a more homogeneous (coeval) sample which  have undergone 
their last
episode of star formation at higher redshift.  
This interpretation is in agreement with our findings in
S\'anchez--Bl\'azquez et al.\ (2003) and in Paper I.

\section{The age--metalicity relation}
\label{sec-age.metallicity.relation}
Several authors have noted that when the age and metallicity obtained from an
index--index diagram are plotted together, they show a correlation in the sense
that younger galaxies seem to be also more metal rich (e.g. Trager et al.\
1998; J\o rgensen 1999; Ferreras et al.\ 1999; Trager et al.\ 2000b; Terlevich
\& Forbes 2002). This age--metallicity relation is difficult to explain under
the hypothesis of passive evolution and high formation ages. 
However, this relation is expected if, during their evolution, galaxies have
undergone several episodes of star formation, in which the new stars formed
from pre-enriched  gas by the previous generations of stars.  Furthermore, the
existence of an age--metallicity relation has implications in the
interpretation of the scale-relations. The low dispersion in the
Mg$_2$--$\sigma$  or the color--magnitude relations, and the existence of a
fundamental plane have been common arguments in favor of the  hypothesis that
elliptical galaxies are old systems which formed all their stars at high
redshift and evolved passively since then (Bender, Burstein \& Faber 1993;
Bernardi et al.\ 1998). However, some authors, (e.g. Trager et al.\ 1998,
2000b; Ferreras et al.\ 1999; J\o rgensen 1999) have studied the scale
relationships showing that a possible age--metallicity degeneracy would
constitute a conspiracy to preserve the low dispersion in those relationships
even when a relatively large fraction of galaxies contain young stars.

\begin{figure}
\centering
\resizebox{1.0\hsize}{!}{\includegraphics[angle=-90]{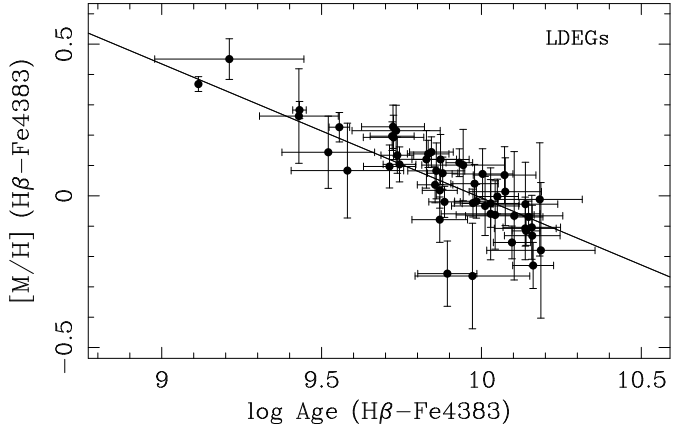}}
  \caption{Age--metallicity relation for LDEGs when the age and metallicity
  are derived from a Fe4383--H$\beta$ diagram. \label{age-metallicity}}
\end{figure}

Fig. \ref{age-metallicity} shows  this correlation for LDEGs when the  age and
metallicity are derived from a Fe4383--H$\beta$ diagram. It is clear from this
diagram that younger galaxies do appear to be  also more metal rich. However,
when  age and metallicity  are measured in a partially degenerated index--index
diagram,
the correlation of the errors in both parameters tend to create an artificial
anti-correlation between them (Kuntschner et al.\ 2001), so it is difficult to
disentangle if the relation is real or an artifact due to the age-metallicity
degeneracy. To check if the correlation of the errors  could be the reason for
the observed trend in our sample, we carried out a similar test to that
performed by Kuntschner et al. (2001). We chose to model three different population
with the following characteristics:
\begin{tabular}{@{}l@{}c@{$\;$}l@{}}
 \raisebox{0ex}[3ex]{\mbox{}}
 (1) &:& \parbox[t]{0.85\columnwidth}{Population with a single age of 8
 Gyr and solar metallicity.}\\
 \raisebox{0ex}[3ex]{\mbox{}}
 (2) &:& \parbox[t]{0.85\columnwidth}{Population with a single age of 8
 Gyr and metallicity ranging from $0.00<{\rm [M/H]}<0.06$.}\\
 \raisebox{0ex}[3ex]{\mbox{}}
 (3) &:& \parbox[t]{0.85\columnwidth}{Population with a range of ages
 between 5.6 and 10 Gyr, and solar metallicity.}
 \raisebox{0ex}[0ex][4ex]{\mbox{}}
\end{tabular}
We measured the Fe4383 and H$\beta$ in these three populations and tried to
reproduce the dispersion due to errors in this index--index diagram through
Montecarlo simulations. To do that, each point was perturbed with our typical
observed error, following a Gaussian distribution. Fig.
\ref{age-meta-simulations} shows the index--index diagrams for 10000 Monte
Carlo realizations of the 3 populations 
(small dots). The model predictions from V06 are also plotted. 
\begin{figure*}
  \centering


  \caption{{\it Top panels}: index--index diagrams for 3 different fake
  populations with the following characteristics: {\it From left to right}, (1)
  Age 8 Gyr and solar metallicity [M/H]=0; (2) Age 8 Gyr and metallicity
  between 0.0$<[$M/H$]<0.06$; (3) Age between 5.6 and 10 Gyr and solar
  metallicity $[$M/H$]$=0. Small dots represent the results of Monte Carlo
  simulations in which each point was perturbed following a Gaussian
  distribution with standard deviation given by the typical error in Fe4383 and
  H$\beta$ indices in the sample of LDEGs. {\it Bottom panels}: Age-metallicity
  relation for each of the 3 fake distributions plotted on the upper panels.
  Dashed lines show the linear fit to these distributions. Grey circles show
  the age and metallicity for LDEGs and the solid lines represent the linear
  fit to these data.\label{age-meta-simulations}}
\end{figure*}
The age and metallicities obtained by interpolating in the diagrams for the
three different populations are represented in the bottom panels of Fig.
\ref{age-meta-simulations} (small dots). The grey circles represent the
observed values in LDEGs.

We then performed a  linear fit to quantify the slope of the relations in both,
the fake distributions and the galaxies. The errors in the slope were estimated
using Monte Carlo simulations in which we generated $N$ elements, ($N$ being
the number of points), where some data appear duplicated and others not. The
final error was obtained as the standard deviation of the slopes in 1000
different simulations. The final results are shown in Table \ref{tab-age-meta}.
We performed a $t$ test to check whether the slopes of the simulations were
compatible with the slope of the LDEGs data. A $t$ value higher than 1.96 
allows us to reject the hypothesis of equal slopes with a significance level
lower than 0.05.  As  can be seen, the slopes defined by the populations 2 and
3 are not compatible with the slope of LDEGs. However, the slope obtained for
the first distribution of constant age and metallicity is marginally compatible
with the one obtained from the data (although the probability that the slopes
are different is $>95$\%).  Nevertheless, none of the three models can
reproduce the dispersion in age and metallicity observed in the sample of
LDEGs. We then conclude that, although the correlation of the errors can
explain the existence of an age--metallicity relation in a distribution of
galaxies with constant age and metallicity, it can not explain the observed
dispersion in these parameters. If we try to reproduce the dispersion  by
simulating populations with a  range in age or metallicity, the slope of the
relation does not reproduce the slope obtained for the  real galaxies.  This
indicates that part of the relation has to be real, and not only a consequence
of the correlation of  errors, although the actual value of the slope can be
modified by this effect. 

\begin{figure}
  \resizebox{1.0\hsize}{!}{\includegraphics[angle=-90]{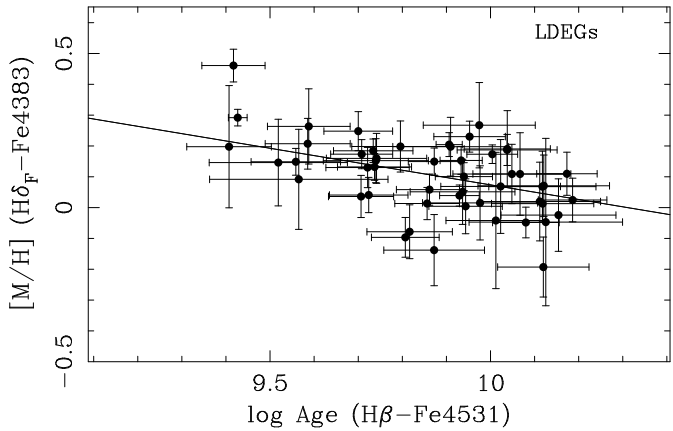}}
  \caption{Comparison of the ages and metallicities obtained from completely
  independent diagrams for the LDEGs. The line represent a least square fits,
  minimizing the residuals in  both directions, x and y.
  \label{fig-edad.metalicidad.independent}} 
\end{figure}

Another way to verify if the relation is an artifact of the error correlations 
is to represent the age versus the  metallicity obtained in two completely
independent index--index diagrams. Fig. \ref{fig-edad.metalicidad.independent}
shows the age--metallicity relation where the ages have been measured in a
H$\beta$--Fe4531 diagram and the metallicities in a H$\delta_F$--Fe4383
diagram.  A non-parametric Spearman rank-order correlation test gives a
correlation coefficient of $-0.47$ corresponding to  a significance level of
0.0002. Although the slope of the relation is flatter ($-0.237\pm 0.076$) than
the one obtained by measuring the ages and metallicities in  a Fe4383--H$\beta$
diagram, there is still a significant correlation, which confirms that  the
age-metallicity relation is not entirely due to the correlation of the errors
in both parameters.
 
The age--metallicity relation for HDEGs is plotted in Fig.
\ref{rel-age.metallicity.coma} (see also Table \ref{tab-age-meta}).  For
comparison, we have also plotted the relation for the LDEGs (dashed line). As
can be seen, the relation for HDEGs seems to be slightly flatter than for the
LDEGs, although a $t$-test does not allow us to discard the possibility that
they are equal within the errors. However, in Paper I we showed that in order
to explain the relation of the indices with $\sigma$ for the HDEGs a variation
of the age along the mass sequence was not necessary.  Thus, if the
age-metallicity relation is a consequence of recent star formation events (in
which the younger stars have been formed from a more enriched gas), we would
not expect to find this age--metallicity relation for the HDEGs.
To verify if the results obtained here are compatible with the relations of the
indices with $\sigma$ reported in Paper I, we carried out the following
experiment: 

\begin{figure}
 \centering
  \resizebox{1.0\hsize}{!}{\includegraphics[angle=-90]{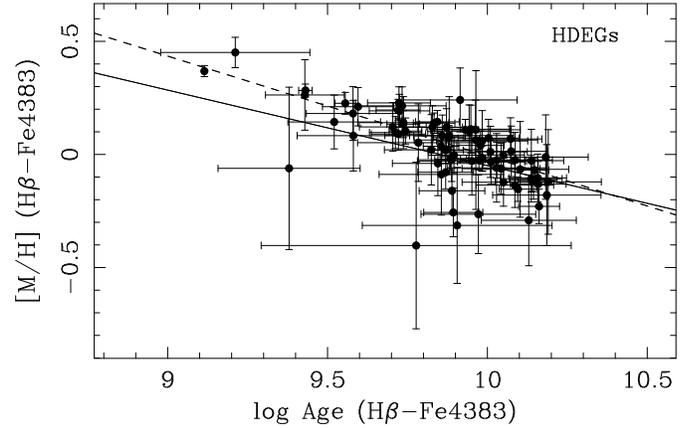}}
  \caption{Age-metallicity relation for HDEGs, where the age and the
  metallicity have been measured in a Fe4383--H$\beta$ diagram. The solid line
  represents a least square fit to the data minimising the residuals in both
  directions.  The dashed line represents the best fit obtained for
  LDEGs.\label{rel-age.metallicity.coma}}
\end{figure}

\begin{table*}
 \caption{Slopes of the age--metallicity relation calculated in the
 Fe4383--H$\beta$ diagram for the three described distributions (see text), and
 for the LDEGs and HDEGs. The last two columns shows the dispersion in age and
 metallicity for the different distributions and the subsamples of observed
 galaxies.\label{tab-age-meta}}
\centering
\begin{tabular}{@{}l r r r rrr@{}}
 \hline\hline
 Distribution & Age(Gyr)   & [M/H] &\multicolumn{1}{c}{slope}& 
 \multicolumn{1}{c}{$t$} & 
 \multicolumn{1}{c}{$\sigma_{\rm \log(age)}$} & 
 \multicolumn{1}{c@{}}{$\sigma_{\rm [M/H]}$}\\
 \hline
 1            & 8.00       & 0.06      & $-0.668\pm 0.005$ & 2.00& 0.084              &  0.073          \\
 2            & 8.00       & 0.00--0.06& $-0.683\pm 0.005$ & 2.37& 0.084              &  0.076         \\
 3            & 5.62--10.00& 0.00      & $-0.362\pm 0.004$ & 5.41& 0.113              &  0.069       \\
  LDEG         &             &           &$-0.516\pm 0.044$&     & 0.255              & 0.152\\
 \hline
  HDEG         &            &            &$-0.333\pm 0.152 $&     & 0.181              & 0.153\\
 \hline
 \end{tabular}
\end{table*}

---  We calculated pairs of indices from the indices-$\sigma$ relations
obtained in Paper I at a variety of $sigma$ sampling from $\sigma=50$km$^{-1}$
to $sigma=300$kms$^{-1}$
in intervals of constant velocity dispersion. Then, we measured the age 
and metallicities of   these fake distribution of indices.
%
We did not try to reproduce the observed distribution,
i.e.  the number of points in each $\sigma$-bin, but just the slope of the
relation.

--- To measured the ages and metallicities of this mock sample we use a
Fe4383--H$\beta$ diagram, obtaining the age--metallicity relation defined by
the index--$\sigma$ relations. The indices for each simulated point were then
perturbed with the observational error adopting a Gaussian probability
distribution, and an age--metallicity relation for the resultant distribution
was also derived.

The age--metallicity relations obtained in this way are plotted in  Fig.
\ref{age.meta.fit}.  We analyse the  results obtained for LDEG and HDEG
separately:

\begin{itemize}
  \item LDEG: The age--metallicity relation obtained for the galaxies following
  the index--$\sigma$ relation has an slope of $0.26\times10^{-5}\pm 10^{-5}$. 
  This
  value is lower than the slope of the real data  but the probability that it
  is significantly different than zero is higher than 99\%. 
  Interestingly, the slope obtained in this way is similar to the slope
  obtained with two independent diagrams ($-0.237\pm 0.076$). We can consider
  this slope to be  more representative of the real slope of the
  age--metallicity relation.

  \item HDEG: For this sample of galaxies, there is no age--metallicity
  relation for the points following the index--$\sigma$ relations.  However,
  when we add the errors, we obtain an artificial age--metallicity relation
  with a slope of $-0.403\pm 6\times 10^{-5}$, compatible, within the errors, 
  with
  the  slope obtained for the galaxies in this subsample. 
\end{itemize}

\begin{figure}
  \resizebox{1.0\hsize}{!}{\includegraphics[angle=-90]{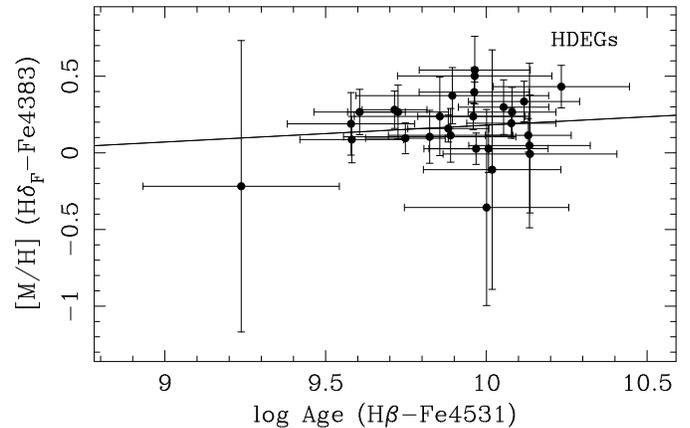}}
  \caption{Age metallicity relation for HDEGs where the age and the metallicity
  have been obtained from two independent diagrams. The line represent a least
  square fits minimazing the residuals in both directions, x and y.
  \label{age.metallicity.independent.coma}} 
\end{figure}

Therefore, we conclude that a real relation between age and metallicity does
exist (i.e. younger galaxies tend to be also more metal rich) for the LDEGs.
On the other hand, HDEGs do not follow this relation, and the age-metallicity
relation shown in Fig. \ref{rel-age.metallicity.coma} is probably a consequence
of the correlation of the errors. In fact, when we measure the age and
metallicity of the HDEGs in two independent diagrams (Fig.
\ref{age.metallicity.independent.coma}) we do not find any correlation between
both parameters (the non-parametric Spearman rank order coefficient is 0.039
with a significance level of 0.422).  The differences in the age--metallicity
relation between LDEGs and HDEGs cannot be a consequence of differences in the
luminosity range of the different samples, because the luminosity coverage of
HDEGs  is somewhat broader than those of LDEGs.  A sample biased toward
high-$\sigma$ galaxies could also make the age-metallicity relation appear
flatter, as there is some evidence that the age dispersion is higher in the
range of  low-$\sigma$ galaxies 
(Poggianti et al.\ 2001b; Caldwell et al.\ 2003).
The sample of HDEGs, however,  is biased  towards low-$\sigma$ galaxies
compared with  the LDEGs (see Paper I).  

\begin{figure}
   \caption{Relation between the age and metallicity obtained for a mock
   distribution of points following the index--$\sigma$ relations (filled
   circles) and for this sample, perturbed with the observational errors (small
   dots). See the text for details.\label{age.meta.fit}}
\end{figure}

\section{Discussion}
In this section we will try to explain all the trends found in the previous
sections with a common scenario.  The results presented in this paper and in
Paper I indicate that HDEGs constitute a more homogenous family than LDEGs;
their stellar populations can be explained under the hypothesis of a single
population and they are, on average, older.  In Fig. \ref{panel}, it has been
shown that this subsample of galaxies exhibits a relation between the
metallicity and the velocity dispersion, no matter which indices are used to
derive this parameter, but, on the contrary, there is not age variation with
velocity dispersion.
For LDEGs, however, the age dispersion is higher and their populations
are best explained as a composition of different bursts of star formation.

The hierarchical clustering models of structure formation
predict different star formation histories
for galaxies situated in different environments (Baugh et al.\ 1996; Kauffmann
\& Charlot 1998; de Lucia et al. 2006). In these models, 
clusters of galaxies are formed from the
highest peaks in the primordial density fluctuations. It is there where the
merging of dark matter haloes, which contained the first galaxies, leads to
galaxies dominated by a bulge at high redshifts ($z\geq2$). The mergers of
galaxies and the adquisition of cold gas cannot continue once the relative
velocity dispersion  between galaxies is higher than 500 km s$^{-1}$, which
makes the occurrence of further star formation episodes in these galaxies more
difficult.  This truncated star formation history also explains the higher
[Mg/Fe] found in HDEGs with respect to the values in younger looking LDEGs
(Paper I). 

On the other hand, the star formation in LDEGs  has probably extended over a
longer period of time, due to the occurrence of more star formation events or
due to a longer single episode of star formation. This scenario was proposed to
explain the differences between N, and maybe C, when comparing galaxies in
different environments (S\'anchez--Bl\'azquez et al.\ 2003, Paper I).

We speculate that  LDEGs and HDEGs could have initially presented similar
relations between the metallicity and the velocity dispersion after their first
massive star formation episode.
However, if LDEGs have suffered subsequent episodes of star formation, the
original correlation between metallicity and potential well (or mass) could
have been erased, 
since other processes could have also  played a role in defining the final
metal content of the galaxies.  The new stars, formed in the more recent
events, would do it from a gas more enriched in the elements produced by low-
and intermediate-mass stars, due to the higher active evolution timescale of
these galaxies.
If these star formation procesess have had a greater relative influence (a
larger ratio between the burst strength and the total galaxy  mass) in less
massive galaxies, as  suggested by the age--$\sigma$ relation, this would
destroy the original relation between mass and metallicity
(increasing Fe, C and N in low velocity dispersion galaxies).
Furthermore, this would result in  a relation between age and metallicity 
(as inferred from Fe features) in LDEGs, but not in HDEGs, as it is found 
in this paper.
Another posibility is that less massive galaxies had actually experienced a
more extended star formation history than more massive galaxies (Chiosi \&
Carraro 2002).  This latter possibility is favoured by some recent studies that
found a depletion in the luminosity function of red galaxies towards the faint
end (Smail et al.\ 2001; de Lucia et al.\ 2004). 

\begin{figure}
  \resizebox{1.0\hsize}{!}{\includegraphics[angle=-90]{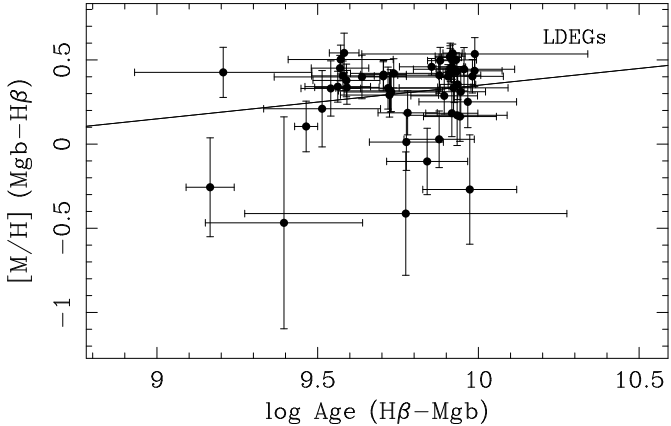}}
  \caption{Age--metallicity relation for the sample of low-density environment
  galaxies  when these parameters are measured in a Mgb--H$\beta$ diagram. The
  line indicate a least square fits to the data, minimizing the residuals in
  both directions x and y. \label{age-metallicity.relation.mgb}}
\end{figure}

Other authors have found differences between the mass--metallicity relation of
galaxies in different environments.
Trager et al.\ (2000b) found that there is a  velocity dispersion--metallicity
relation  for old cluster galaxies, but no comparable relation exists for field
ellipticals.  This result is compatible with ours,  with the difference that we
still find a steep relation between the metallicity and the velocity dispersion
for LDEGs when the metallicity is measured with Mgb. Actually, Trager et al.\
also found a relation between what they called the {\it enhanced\/} elements
(including Mg) and velocity dispersion for all the galaxies in their sample.

If, as we have argued, the age--metallicity relation is a consequence of later
episodes of star formation, and the relative enrichment have been more
pronounced in the Fe-peak elements,  we would expect differences in 
the age-metallicity relation when the metallicity is measured using 
an index with a different sensitivy to changes in Fe and Mg. 
. Fig. \ref{age-metallicity.relation.mgb}
shows the age--metallicity relation when these parameters are measured in a Mgb--H$\beta$
diagram. The non-parametric rank order coefficient is 0.177 with a significance level of
0.10. Certainly, there is not a significant correlation between these two
parameters when the Mgb index is used instead of Fe4383.  We need to 
stress again that we are not calculating chemical abundances in this paper. The
metallicity measure with Mgb does not correspond to  the abundance of Mg, and neither the metallicity
measured with Fe4383, an Fe abundance. We argue though, that the different behaviours of
the metallicities calculated with different indices are the consecuence of their
different sensitivities to variation of different chemical species. In this specific case, 
the flatter slope of the age-metallicity relation when a more (less) sensitive Mg (Fe)
index is used is  in agreement with our
scenario.

Interestingly, the more massive galaxies in low density environments show a
behaviour very similar to the massives galaxies of the Coma cluster.  These
very massive galaxies tend to have boxy isophotes, which can be explained by
models of mergers without gas (Binney \& Petrou 1985; Bender \& M\"{o}llenhoff
1987; Nieto \& Bender 1989; Nieto et al. 1991; Bender, Burstein \&
Faber 1992; Faber et al.\ 1997; Lauer et al.\ 2005), since a few percent of the
mass in gas is sufficient to destroy boxy orbits and impart high global
rotation (Barnes 1996; Barnes \& Hernquist 1996).  Furthermore, boxy galaxies
tend to have {\it cores} inner profiles (Faber et al. 1997). N-body simulations
of merging galaxies with central black  holes (Ebisuzaki, Makino \& Okumara
1991; Makino 1997; Quinlan \& Hernquist 1997; Milosavljevi\'c \& Merritt 2001)
show that cores can indeed form in such merger remnants. Recently, Lauer et
al.\ (2005) have found  that power-law galaxies, on average, have steeper
colour gradients than do core galaxies (although the difference is small). This
result is compatible with the idea that power-law galaxies have formed in
gas-rich mergers while core galaxies have formed from free-gas mergers, which
would cause a dilution in the metallicity gradient.  Actually, these mergers
without gas have been observed in clusters at $z=0.8$ (van Dokkum et al.\
1999).  The existence  of these gas-free mergers indicates that the epoch of
assembly does not necessarily coincide with the epoch of formation of the bulk
of stars.  This scenario could bring into agreement the hierarchical models of
galaxy formation with the observed trends of age with mass for elliptical
galaxies in LDEGs. These trends (low-$\sigma$ galaxies appearing to be younger)
are completely opposite to what is expected under  these scenarios of galaxy
formation, that predict that larger galaxies assemble at later times than small
ones (Kauffman, White \& Guiderdoni  1993). However, these predictions are made under the
assumption that all the gas cooled off and formed stars when the haloes were
assembled.  However,  other processes such as supernova feedback may play a
role in regulating the rate at which stars form in these systems (e.g. Kawata
\& Gibson 2003).   Several mechanisms have been proposed to explain the
appearence that low-mass galaxies have suffered a more extended star formation
history. Kawata (2001) suggests that UV background radiation is a possible
candidate, because it suppresses cooling and star formation more strongly in
lower mass systems (Efstathiou 1992), and is expected to extend the duration of
star formation. Chiosi \& Carraro (2002) have recently build N-body-tree-SPH
simulations incorporating cooling, star formation, energy feedback and chemical
evolution. These authors find that the star formation history is governed by
the initial density and total mass of the galaxy, and that the interplay of the
above processes results in a more extended star formation history in low-mass
galaxies.  Until we  understand completely the role of these mechanisms, we
will not be able to rule out different processes of galaxy formation. 

 \section{Conclusions}
 
We have studied the stellar population properties of the centers of  98
early-type galaxies spaning a  large range in velocity dispersion. Using the
new stellar population synthesis models of V06, which include a new and
improved stellar library (MILES), we have derived ages and metallicities for
this sample of galaxies. Due to the difficulties in deriving chemical
abundances with the available tools, we have studied the behaviour of the
different chemical elements in a very qualitative way, measuring the
metallicity with different indices especially sensitive to different chemical
species. From this analysis, we conclude: 

\begin{itemize} 
  \item The sample of LDEGs spans a wide range in SSP-equivalent ages and
  metallicities. This confirms previous results obtained by other authors (e.g.
  Gonz\'alez 1993; Trager et al.\ 1998; Trager et al.\ 2000b). This age spread
  is not meant to imply that galaxies formed all their stars at different
  epochs. In fact, we have shown that galaxies in low density environments are
  best explained as a composition of populations in which a low percentage of
  young stars is added to an old population, in agreement with the conclusions
  of Trager et al.\ (2000b).

  \item For the subsample of LDEGs, there is a relation between the age and the
  velocity dispersion in the sense that less massive galaxies tend to be
  younger.  This ``down-sizing effect'' suggests either that the episodes of
  star formation have had a larger relative influence on the low mass galaxies,
  or that the last star formation activity occurs on average at lower redshifts
  for progressively fainter galaxies.   This relation is also present  
  for   
  galaxies in the Virgo cluster,  but it is not present  in the
  subsample of galaxies in the Coma cluster.
 
  \item Comparing the ages obtained in different regions of the spectra we have
  shown that galaxies in low-density environments are best described by a
  composition of populations in which an small percentage of young stars are
  added to an old  population. On the other hand, the population of the HDEGs
  can be described with a single burst, which does not necessarily indicate
  that these galaxies formed all their stars in a  single burst, but may
  indicate that the last episode of star formation finished at earlier times
  than in LDEGs.

  \item The sample of LDEGs shows a relation between the age and the
  metallicity implying that younger galaxies are also more metal rich. This
  relation is true even when the age and metallicity are measured in completely
  independent index--index diagrams, indicating that it is not a consequence of
  a correlation of the errors in both parameters. However, the actual relation
  can be flatter than the relation derived from a partially degenerate
  index--index diagram. We have shown that this relation is only evident when
  some indicators are used to measure the metallicity. In particular, we do not
  find a relation when the metallicity is measured in a  Mgb--H$\beta$ diagram.
  If the age--metallicity relation is a consequence of the occurrence of late
  star formation in these galaxies, this would imply that the relative
  enrichment in Mg in the last generations of stars is much less important with
  respect to the Fe.
  The sample of HDEGs,  however, do not show this relation between the age and
  the metallicity, which is in agreement with the general picture exposed in
  this work (and in Paper I) in which HDEGs have had a truncated star formation
  history compared to their counterparts in low-density environments. 

  \item There exists a mass--metallicity relation for HDEGs, in the sense
  that more massive galaxies tend to be also more metal rich. This is
  independent of the  indicator used to measure the metallicity, although the
  slope is somewhat swallower if Fe4383 is used instead of Mgb or CN$_2$.
  However, in the case of the LDEGs the mass--metallicity relation is only
  apparent  when the metallicity is measured with Mgb. When the metallicity is
  measured with Fe4383 or CN$_2$, younger galaxies tend to lie at higher
  metallicities for a given $\sigma$. This can indicate that the
  mass--metallicity relation is a consequence of processes that occurred when
  the bulk of the stars were formed. In LDEGs, later star formation events
  raise the metallicity in low-mass systems, flattening this relation. As the
  relative enrichment in these events is more pronounced in the elements
  produced by low-mass stars, the flattening in the relation, when the
  metallicity is measured with indices sensitives to these elements, is also
  more evident.

\end{itemize}

Our results show that there exist differences in the stellar populations of
galaxies inhabitating different environments. HDEGs represent a more homogenous
sample of galaxies than LDEGs, their stellar populations can be explained
assuming a single burst and their mean ages are slighly higher.  On the other
hand, LDEGs show more variety in their stellar populations. They span a large
range of ages and their spectral features are better explained asuming that
they have suffered multiple bursts of star formation.

It is worth recalling that the results discussed in  Papers I and II of this
series refer only to the central regions of the galaxies. If significant radial
age and metallicity gradients are present within the galaxies, these results
cannot be considered representative of the whole star formation history. That
is,  to constrain the star formation history of early-type galaxies, we cannot
ignore the behaviour of the SSP-parameters along the radii.  This analysis is
beyond the scope of the present paper, but will be the subject of the third
paper of the series.

\section*{Acknowledgments}
We are very grateful to the referee, Jim Rose, for his very constructive report and many
useful suggestions. We are also grateful
to Javier Cenarro, Reynier Peletier and Alexandre Vazdekis for many fruitful discussions.
The WHT is
operated on the island of La Palma by the Royal Greenwich Observatory at the
Observatorio del Roque de los Muchachos of the Instituto de Astrof\'{\i}sica de
Canarias.
The Calar Alto Observatory is operated jointly by the Max-Planck Insitute
f\"{u}r Astronomie, Heidelberg, and the Spanish Instituto de Astrof\'{\i}sica
de Andaluc\'{\i}a (CSIC). This work was  supported by the Spanish research
project AYA 2003-01840 and by the Australian Research Council. We are grateful
the CAT for generous allocation of telescope time.

\label{lastpage}
\end{document}